\documentclass[10pt,letterpaper]{article}

\usepackage[utf8]{inputenc}
\usepackage[T1]{fontenc}
\usepackage{newtxtext}
\usepackage[letterpaper,textwidth=6.5in,textheight=9in,top=1in,footskip=30pt]{geometry}
\usepackage{amsmath,amssymb,bm,mathtools}
\usepackage{microtype}
\usepackage{xcolor}
\usepackage{graphicx}
\usepackage{booktabs}
\usepackage{rotating}
\usepackage{placeins}
\usepackage[hidelinks]{hyperref}

\makeatletter
\renewcommand\section{\@startsection{section}{1}{\z@}%
  {-2.0ex \@plus -0.5ex \@minus -0.2ex}{1.5ex \@plus 0.3ex}{\large\bfseries}}
\renewcommand\subsection{\@startsection{subsection}{2}{\z@}%
  {-1.8ex \@plus -0.5ex \@minus -0.2ex}{0.8ex \@plus 0.2ex}{\normalsize\bfseries}}
\renewcommand\subsubsection{\@startsection{subsubsection}{3}{\z@}%
  {-1.5ex \@plus -0.5ex \@minus -0.2ex}{0.5ex \@plus 0.2ex}{\normalsize\bfseries}}

\renewcommand{\@maketitle}{%
  \begin{center}
    {\LARGE\scshape \@title\par}
    \vskip 0.3in
    \@author
  \end{center}
  \vskip 0.15in}
\makeatother

\newcommand{\dd}{\,\mathrm{d}}

\newcommand{\vect}[1]{\bm{#1}}

\newcommand{\clip}{\operatorname{clip}}

\title{Unified Projection-Domain Material Decomposition\\for Energy-Integrating and Photon-Counting CT}
\author{%
  \textbf{Dale Black}$^{1}$, \textbf{Surya Naidu}$^{1}$,
  \textbf{Hamidreza Khodajou-Chokami}$^{1}$, and \textbf{Sabee Molloi}$^{1,*}$\\[1.5ex]
  $^{1}$Department of Radiological Sciences, University of California, Irvine, CA 92697, USA\\[1ex]
  $^{*}$Corresponding author: \texttt{symolloi@uci.edu}}
\date{}

\begin{document}

\maketitle

\begin{center}
  {\large\scshape Abstract}
\end{center}
\begin{quote}
\sloppy

\noindent\textbf{Objective:}
To develop a unified and efficient {method} for
projection-domain material decomposition in spectral computed tomography (CT) that
incorporates an arbitrary number of spectral measurements, independent of
the acquisition mechanism, without {reducing the measurements}
to two effective {energies}.

\medskip
\noindent\textbf{Approach:}
A nested optimization {method} was developed for two-material basis
decomposition from $K\geq2$ spectral measurements. It extends an existing
two-measurement univariate method by replacing its polynomial approximation
with profile {minimization of the Poisson negative log-likelihood}.
The two-variable inversion was reduced to two one-dimensional searches that use all
measurements without spectral rebinning. The {method} was evaluated {in simulation} for energy-integrating (GE Revolution Apex Elite,
Siemens SOMATOM Flash) and photon-counting (Siemens NAEOTOM Alpha) CT. Estimated line integrals were compared with the truth and
the Cram\'er--Rao lower bound (CRLB), and CT-number accuracy in Hounsfield
units (HU), image noise, speed, and numerical robustness were evaluated.

\medskip
\noindent\textbf{Main results:}
The {method} recovered line integrals with
noiseless iodine errors below $3\times10^{-6}$~g\,cm$^{-2}$ and noise within
3\% of the CRLB. Virtual monoenergetic images (VMIs) generated from them
agreed{, at 70~keV and above,} with reference CT numbers and{, on all three systems,} with {vendor VMIs of} physical scans
(mean absolute error 6--22~HU). For the photon-counting acquisition,
using four measurements instead of two reduced the
iodine CRLB by 19\% and image noise by 8--16\% at every VMI energy,
and lowered the CT-number regression error at 70~keV and above (3.1 versus
5.2~HU at 70~keV). {Iteration and tolerance} settings changed
CT numbers by less than 0.04~HU, and decomposition was fast (0.34--0.67~s
per million rays on an A100 GPU).

\medskip
\noindent\textbf{Significance:}
Using low-dimensional optimization and all measurements, the
{method} provides a fast, accurate, and generalized decomposition for any
spectral CT acquisition mechanism and number of measurements. One method can
serve every scanner in a virtual imaging trial.


\smallskip
\noindent\textbf{\textit{Abbreviations}}\quad {CPU, central processing unit;} CRLB, Cram\'er--Rao lower bound; CT, computed tomography; EICT, energy-integrating detector CT; FBP, filtered backprojection; {GPU, graphics processing unit;} HU, Hounsfield unit; MAE, mean absolute error; PCCT, photon-counting detector CT; ROI, region of interest; SD, standard deviation; VIT, virtual imaging trial; VMI, virtual monoenergetic image.
\end{quote}

\section{Introduction}

Spectral computed tomography (CT) uses the energy dependence of x-ray attenuation to provide quantitative information beyond that available from conventional single-energy CT. {Measurements at two or more effective energies allow the total attenuation to be separated into material-specific components.} This capability forms the foundation of several spectral imaging applications, including material-specific imaging, virtual monoenergetic imaging, virtual noncontrast imaging, and K-edge imaging~\cite{WangPelc2021,McCollough2015,McCollough2023,Leng2019}. The term spectral CT covers two detector technologies that acquire these measurements in different ways. Energy-integrating detector CT (EICT) records the total energy deposited by the detected photons{. Its effective energies come either from changing the x-ray spectrum, as in rapid kVp switching, dual-source acquisition, and split filtration, or from a dual-layer detector, whose stacked layers absorb the lower- and higher-energy parts of a single spectrum}~\cite{McCollough2015,McCollough2020}. Photon-counting detector CT (PCCT) instead counts individual photons and sorts them by energy into two or more energy {windows} within a single acquisition~\cite{Leng2019,WangPelc2021}. Systems of both kinds are increasingly studied in virtual imaging trials (VITs), in which CT simulators {model} the acquisition of specific vendor systems so that they can be evaluated on the same virtual phantoms and patients~\cite{Abadi2020,AbadiDukeSim2019,AbadiPCCT2019,Black2026}.

Material decomposition is the fundamental inverse problem underlying these applications. Within the diagnostic x-ray energy range, the attenuation of most biological tissues can be represented using a small number of energy-dependent basis functions~\cite{Alvarez1976}. In two-material basis decomposition, the spectral measurements are used to estimate the line integrals of two basis materials, such as water and iodine. The resulting basis maps can be combined into material-specific maps or into virtual monoenergetic images (VMIs), which approximate the image that a monoenergetic x-ray beam would produce and are widely used clinically~\cite{McCollough2015}. A two-material basis is generally sufficient for mixtures of low-atomic-number biological tissues, whereas additional basis functions may be required when K-edge contrast materials are present~\cite{Tang2021}. The feasibility of quantitative tissue characterization using spectral CT has been demonstrated in several applications, including the decomposition of breast tissue into water, lipid, and protein components~\cite{Ding2012,Ding2014}.

Material decomposition methods are commonly classified as image-domain, projection-domain, or one-step approaches. Image-domain methods decompose independently reconstructed spectral images and are straightforward to implement, but because each image is reconstructed from polychromatic data, they often retain beam-hardening artifacts and quantitative bias unless additional corrections are applied~\cite{Maass2009,Maass2011}. When measurements along the same ray paths are available, projection-domain methods can estimate the basis line integrals directly from them before reconstruction, thereby reducing beam-hardening artifacts. {Such measurements are available in PCCT and dual-layer CT, and in rapid kVp switching after the alternating views are interpolated onto common ray paths~\cite{McCollough2015}.} These methods {often} require accurate spectral and detector models or calibration measurements~\cite{Andriiashen2021,Alvarez2011,Rajbhandary2017}. One-step methods solve decomposition and reconstruction as a single problem: an iterative reconstruction estimates the basis images directly from the measured spectral projections, without first forming basis line integrals or spectral images. This joint formulation can incorporate spatial regularization, but such methods are generally computationally demanding and introduce regularization or constraint parameters that must be tuned~\cite{Schmidt2017,Long2014,Tilley2019,Mory2018}. More recently, deep-learning-based methods have been investigated to improve decomposition accuracy, suppress noise, and reduce material cross-talk{. However,} these methods generally require representative training data, which can be difficult to obtain clinically~\cite{Gong2020,Nadkarni2022,Peng2024}. Projection-domain decomposition therefore provides a practical balance: like one-step methods, it models the polychromatic physics and thus reduces beam hardening, yet in its standard form it decomposes each ray independently, without spatial regularization, and its basis sinograms can be reconstructed and denoised with conventional methods.

The fundamental problem in projection-domain material decomposition is the inversion of the nonlinear polychromatic forward model. Conventional approaches commonly estimate the basis line integrals using multidimensional iterative optimization~\cite{Ducros2017}, numerical lookup tables~\cite{Zimmerman2015}, or calibration-based polynomial mappings~\cite{Naidu2026}. Multidimensional optimization requires numerous forward-model evaluations and can be sensitive to initialization and algorithmic parameter selection~\cite{Alvarez2011,Ducros2017,Rajbhandary2017,Cong2022}. Lookup-table and empirical calibration methods can provide rapid inversion but require sufficiently dense calibration data and may not generalize beyond the acquisition and detector conditions represented during calibration.

Cong et al.~\cite{Cong2022} proposed an efficient projection-domain method for two-material basis decomposition in dual-energy CT, the two-measurement configuration of most clinical EICT systems. Their method uses an odd-order polynomial approximation to express one basis line integral as a function of the other, thereby reducing the original two-variable nonlinear inversion to a univariate optimization problem. Because it inverts a physical forward model directly, the method also requires no regularization parameters, training data, or calibration tables, only knowledge of the spectral responses. However, the method is formulated specifically for two measurements acquired with two x-ray spectra, as in dual-kVp CT, and for two unknown basis line integrals. Consequently, it does not directly provide an estimator for a system containing more than two spectral measurements. Such systems arise naturally in PCCT, in which detected photons are sorted into multiple energy {windows}, whereas clinical EICT systems typically provide two spectral measurements. Application of a dual-energy decomposition {method} to PCCT data {therefore} requires the original measurements to be reduced to two effective measurements, for example by summing the energy {windows} into low- and high-energy measurements. Although this rebinning permits the use of an existing dual-energy method, it discards spectral information from the original measurements and makes the decomposition dependent on {how the energy windows are merged}.

In PCCT, rebinning is often motivated by the lower number of photons in each narrow energy {window}. Splitting the detected photons into more {windows} does lower the signal-to-noise ratio of each individual measurement~\cite{Taguchi2024}. Material decomposition, however, uses all of the measurements together, and narrower {windows} sample the energy dependence of the attenuation more finely. Summing {windows} discards part of that information and cannot add to it. {The decomposition noise of any unbiased estimator is bounded from below by} the Cram\'er--Rao lower bound (CRLB). For a simulated photon-counting detector with charge sharing and electronic noise, Taguchi~\cite{Taguchi2024} showed that the CRLB decreased steadily as energy {windows} were added, {albeit} with diminishing returns.

In this work, we propose a nested optimization {method} for two-material basis decomposition from an arbitrary number of spectral measurements. The method builds directly on the univariate decomposition of Cong et al.~\cite{Cong2022}: it retains the reduction to one-dimensional searches but replaces the polynomial relation between the two basis line integrals with a profile minimization of the {negative log-likelihood} over all measurements. The resulting {reformulation as} two coordinated one-dimensional searches is exact, involves no series approximation, and requires no spectral rebinning. Because every measurement enters through the Poisson likelihood, {which weights each by the inverse of its expected count}, the estimate can also approach the CRLB. For two measurements {that can be fit exactly}, both methods seek the same solution{. For} more than two, every measurement is used directly. Each measurement is also described by its own effective spectral response, which includes the detector response, so measurements separated by the detector, such as PCCT energy {windows}, enter the decomposition in the same way as measurements separated by the x-ray spectrum. Like the method of Cong et al.~\cite{Cong2022}, the proposed {method} requires no regularization parameters, training data, or calibration tables. In addition, the proposed estimator is built into an open-source CT simulator, \texttt{BasisSimulator.jl}~\cite{Black2026}, together with the univariate method, rather than provided as a standalone algorithm. Any acquisition the simulator models can therefore be decomposed without reimplementation, which makes the method directly usable in VITs. %

The proposed {method} is applicable to both EICT, with two or more spectral measurements, and PCCT, provided that the measurement model is identifiable and the spectral response functions contain sufficient independent information for estimating the selected basis maps. This generality is particularly useful for VITs, in which several vendor-specific acquisitions are simulated. Applying one decomposition method to every simulated acquisition keeps differences between the scanners from being confounded with differences between their decomposition algorithms. The method is evaluated on three distinct types of spectral CT acquisition: two dual-energy EICT acquisitions, rapid kVp switching (GE Revolution Apex Elite) and dual-source (Siemens SOMATOM Definition Flash), and a PCCT acquisition with four energy windows (Siemens NAEOTOM Alpha). Each is simulated with noiseless and noisy measurements, and the simulated results are compared with physical scans of the same phantom on all three systems. The decomposition is evaluated directly on the estimated line integrals and through VMIs synthesized from them. Decomposition accuracy, noise propagation, convergence behavior, and computational efficiency are compared with those of the univariate method of Cong et al.~\cite{Cong2022}.

The remainder of the paper is organized as follows. Section~\ref{sec:methods} describes the proposed {method} and evaluation methodology. Section~\ref{sec:results} presents the results. Section~\ref{sec:discussion} discusses the performance and limitations of the proposed method. Section~\ref{sec:conclusion} concludes
the paper.

\section{Methodology}
\label{sec:methods}

\subsection{Two-material basis measurement model}
\label{sec:forward_model}

Within the diagnostic energy range, the energy- and space-dependent linear
attenuation coefficient is represented using two energy-dependent basis
functions~\cite{Alvarez1976,Lehmann1981},
\begin{equation}
\label{eq:basis_model}
\mu(\vect r,E)
=
\rho_1(\vect r)f_1(E)
+
\rho_2(\vect r)f_2(E),
\end{equation}
where $f_1(E)$ and $f_2(E)$ are the known mass-attenuation coefficients of
the selected basis materials, and $\rho_1(\vect r)$ and $\rho_2(\vect r)$
are the corresponding spatial density maps. For a single x-ray path $\ell$,
the corresponding basis line integrals are
\begin{equation}
\label{eq:line-integrals}
A=\int_\ell \rho_1(\vect r)\dd\ell,
\qquad
C=\int_\ell \rho_2(\vect r)\dd\ell.
\end{equation}
In this work the two basis materials are iodine and water, so that $A$ and
$C$ are the iodine and water line integrals.

For $K\geq2$ spectral measurements, the expected value of measurement
$k=1,\ldots,K$ is
\begin{equation}
\lambda_k(A,C)
=
\int_{0}^{E_{\max}}
S_k(E)
\exp\!\left[-f_1(E)A-f_2(E)C\right]\dd E,
\label{eq:continuous_forward}
\end{equation}
where $S_k(E)$ {is the spectral response of measurement $k$, the
expected number of photons detected per unit energy in the absence of the
object (in count-equivalent units for an energy-integrating detector),} including the incident spectrum, filtration, exposure, acquisition
geometry, and energy-dependent detector response. The acquisition mechanism
is therefore represented through $S_k(E)$ without changing the form of the
forward model. {The upper limit $E_{\max}$ is the maximum photon energy, set by the
tube potential.}

For numerical implementation, Eq.~\eqref{eq:continuous_forward} is
discretized over $N_E$ energy samples as
\begin{align}
\psi_e(A,C)
&=
\exp\!\left(-f_{1e}A-f_{2e}C\right),
\label{eq:transmission}\\
\lambda_k(A,C)
&=
\sum_{e=1}^{N_E}q_{ke}\psi_e(A,C),
\label{eq:discrete_forward}
\end{align}
where $f_{1e}=f_1(E_e)$, $f_{2e}=f_2(E_e)$, and $q_{ke}$ is the expected
unattenuated contribution of energy sample $e$ to measurement $k$, including
the energy-sampling interval. The responses $q_{ke}$ are specified for each detector ray, so a spectrum that
varies across the detector, such as that produced by a bowtie filter, requires
no change to the estimator. The complete expected measurement vector is
\begin{equation}
\boldsymbol{\lambda}(A,C)
=
\left[
\lambda_1(A,C),\ldots,\lambda_K(A,C)
\right]^{\mathsf T}.
\label{eq:measurement_vector}
\end{equation}

\subsection{Statistical model and objective function}
\label{sec:likelihood}

After {scatter correction, and with the remaining detector effects included in $S_k(E)$,} the measured counts are modeled as conditionally independent Poisson random variables,
\begin{equation}
Y_k\mid A,C\sim\operatorname{Poisson}\!\left[\lambda_k(A,C)\right],
\qquad k=1,\ldots,K.
\label{eq:poisson_model}
\end{equation}
For ideal nonoverlapping PCCT energy windows, this model follows from Poisson splitting. Residual correlations caused by charge sharing, pulse {pile-up}, cross-talk, or imperfect corrections are not represented by Eq.~\eqref{eq:poisson_model} and are treated as model mismatch. For energy-integrating detectors, $Y_k$ is the measured signal expressed in count-equivalent units, and Eq.~\eqref{eq:poisson_model} approximates its compound-Poisson statistics~\cite{Whiting2006}. Additive electronic noise would add a Gaussian term to this model~\cite{Whiting2006,XuTsui2009}. Its variance is small compared with the quantum noise except at very low signal levels, such as along the most highly attenuated rays~\cite{Whiting2006,XuTsui2009}, so it is not modeled separately here and is treated as model mismatch.

Denoting by $y_k$ the observed value of $Y_k$ and ignoring terms independent of $A$ and $C$, the negative log-likelihood is
\begin{equation}
\mathcal{L}(A,C)
=\sum_{k=1}^{K}
\left[\lambda_k(A,C)-y_k\log\lambda_k(A,C)\right].
\label{eq:negative_log_likelihood}
\end{equation}
The joint material-decomposition estimate is defined by
\begin{equation}
(\widehat A,\widehat C)
=\underset{(A,C)\in\Theta}{\arg\min}\;\mathcal{L}(A,C),
\label{eq:joint_estimator}
\end{equation}
where $\Theta=[A_{\min},A_{\max}]\times[C_{\min},C_{\max}]$ is a bounded feasible box whose limits are given in Eq.~\eqref{eq:numerical-controls}. The bounds keep both one-dimensional searches finite.

\subsection{Gradient and Fisher information}
\label{sec:fisher}

The updates of Section~\ref{sec:nested_optimization} use the gradient of
Eq.~\eqref{eq:negative_log_likelihood} and the Fisher information as its
{expected} curvature. With the derivatives of the expected counts
\begin{equation}
\lambda_{k,A}\equiv\frac{\partial\lambda_k}{\partial A}
=-\sum_{e=1}^{N_E}q_{ke}f_{1e}\psi_e(A,C),
\qquad
\lambda_{k,C}\equiv\frac{\partial\lambda_k}{\partial C}
=-\sum_{e=1}^{N_E}q_{ke}f_{2e}\psi_e(A,C),
\label{eq:lambda_A}
\end{equation}
the components of the gradient are
\begin{equation}
g_A(A,C)=\sum_{k=1}^{K}\left(1-\frac{y_k}{\lambda_k}\right)\lambda_{k,A},
\qquad
g_C(A,C)=\sum_{k=1}^{K}\left(1-\frac{y_k}{\lambda_k}\right)\lambda_{k,C},
\label{eq:score_A}
\end{equation}
and, for the independent-Poisson model, the Fisher information matrix is
\begin{equation}
\begin{gathered}
\boldsymbol{F}(A,C)
=\begin{bmatrix}
F_{AA} & F_{AC}\\
F_{AC} & F_{CC}
\end{bmatrix},\\
F_{AA}=\sum_{k=1}^{K}\frac{\lambda_{k,A}^{2}}{\lambda_k},\quad
F_{AC}=\sum_{k=1}^{K}\frac{\lambda_{k,A}\lambda_{k,C}}{\lambda_k},\quad
F_{CC}=\sum_{k=1}^{K}\frac{\lambda_{k,C}^{2}}{\lambda_k}.
\end{gathered}
\label{eq:fisher_information}
\end{equation}
Evaluated at the true $(A,C)$, its inverse is the CRLB, {a lower bound on the covariance
of any unbiased estimator under the usual regularity conditions}~\cite{Alvarez2011,Taguchi2024}. {CRLB values are
reported as standard deviations, the square roots of its diagonal elements.} It is
the reference for the efficiency analysis of Section~\ref{sec:eval_metrics}.

\subsection{Initialization}
\label{sec:initialization}

An approximate log-linear solution is used to initialize the nonlinear estimator. The expected air scan for measurement $k$ is
\begin{equation}
I_{0k}=\sum_{e=1}^{N_E}q_{ke},
\label{eq:air_count}
\end{equation}
and the stabilized log measurement is
\begin{equation}
p_k=-\log\!\left[\frac{\max(y_k,\varepsilon)}{I_{0k}}\right],
\label{eq:log_measurement}
\end{equation}
where $\varepsilon>0$ prevents evaluation of the logarithm at zero. 
Using the normalized air-spectrum weights
\begin{equation}
w_{ke}^{(0)}=\frac{q_{ke}}{\sum_{e'=1}^{N_E}q_{ke'}},
\label{eq:air_weights}
\end{equation}
the effective attenuation coefficients are calculated as
\begin{equation}
\overline f_{1k}=\sum_e w_{ke}^{(0)}f_{1e},
\qquad
\overline f_{2k}=\sum_e w_{ke}^{(0)}f_{2e}.
\label{eq:effective_coefficients}
\end{equation}

The initial basis estimate is then obtained from the linear least-squares problem
\begin{equation}
\boldsymbol{x}^{(0)}
=\underset{\boldsymbol{x}}{\arg\min}\;
\left\|\boldsymbol{M}\boldsymbol{x}-\boldsymbol{p}\right\|_2^2,
\qquad
\boldsymbol{x}=\begin{bmatrix}A&C\end{bmatrix}^{\mathsf T},
\label{eq:linear_initializer}
\end{equation}
where row $k$ of $\boldsymbol{M}$ is $[\overline f_{1k},\overline f_{2k}]$. The resulting $2\times2$ normal equations use all $K$ measurements and are solved in closed form. If they are singular (determinant at most $10^{-12}$), the initial estimate is set to $(A,C)=(0,20)$~g\,cm$^{-2}$. Because the effective-energy approximation does not model beam hardening, this estimate is used only for initialization and is projected onto $\Theta$ before nonlinear optimization.

The initial value of $C$ is then refined by total-count matching. At fixed $A^{(0)}$, define
\begin{equation}
\Lambda(A^{(0)},C)=\sum_{k=1}^{K}\lambda_k(A^{(0)},C),
\qquad
y_{\mathrm{tot}}=\sum_{k=1}^{K}y_k.
\label{eq:total_counts}
\end{equation}
When $f_2(E)>0$ over the contributing energy range, $\Lambda(A^{(0)},C)$ is strictly decreasing in $C$. Therefore, if the residual $\Lambda(A^{(0)},C)-y_{\mathrm{tot}}$ changes sign over the feasible interval, the count-matching equation
\begin{equation}
\Lambda(A^{(0)},C)=y_{\mathrm{tot}}
\label{eq:bisection_initializer}
\end{equation}
has a unique solution within that interval and is solved by bisection. If no valid bracket exists, the least-squares estimate is retained. This refinement is used only to initialize the likelihood optimization and is not imposed as a condition on the final estimate.

\subsection{Nested optimization}
\label{sec:nested_optimization}

For each feasible value of $A$, define the conditional estimate $C^{\star}(A)$, the value of $C$ that minimizes the objective at that $A$, as
\begin{equation}
C^{\star}(A)
\in\underset{C\in[C_{\min},C_{\max}]}{\arg\min}\;\mathcal{L}(A,C),
\label{eq:inner_problem}
\end{equation}
and the profile objective as
\begin{equation}
\mathcal{L}_{\mathrm{p}}(A)
=\mathcal{L}\!\left[A,C^{\star}(A)\right].
\label{eq:profile_objective}
\end{equation}
The joint estimator in Eq.~\eqref{eq:joint_estimator} can consequently be written as
\begin{equation}
\widehat A
=\underset{A\in[A_{\min},A_{\max}]}{\arg\min}\;\mathcal{L}_{\mathrm{p}}(A),
\qquad
\widehat C=C^{\star}(\widehat A).
\label{eq:outer_problem}
\end{equation}

Because the feasible set is a Cartesian product,
\begin{equation}
\label{eq:iterated-minimum}
\min_{(A,C)\in\Theta}\mathcal L(A,C)
=
\min_{A\in[A_{\min},A_{\max}]}
\left[
\min_{C\in[C_{\min},C_{\max}]}\mathcal L(A,C)
\right].
\end{equation}

This formulation is equivalent to the original two-parameter optimization when the inner problem is solved globally for every {feasible} value of $A$. The formulation reduces each search to one dimension but does not, by itself, guarantee convexity of either objective.

The construction parallels the univariate method of Cong et al.~\cite{Cong2022}. Its contribution was to reduce the two-variable projection-domain inversion of dual-energy CT to a single one-dimensional search, by {expressing one basis line integral as a function of the other through an odd-order polynomial equation}. The profile minimization keeps that reduction to one-dimensional searches, but it obtains the relation between the two line integrals, $C^{\star}(A)$, from the likelihood itself at each $A$ rather than from a series approximation. Because the likelihood sums over all measurements, the same construction applies to any number $K\geq2$ of spectral measurements. Cong et al.~\cite{Cong2022} also formulated their method for two measurements acquired with two x-ray spectra{. Here} each measurement has its own spectral response $S_k(E)$ (Eq.~\eqref{eq:continuous_forward}), which also contains the detector response{. Detector-separated measurements are therefore} handled in the same way as source-separated ones.

\subsubsection{Inner optimization}

At fixed $A$, the inner problem is solved using a projected Fisher-scoring update,
\begin{equation}
\label{eq:inner-update}
C^{(t+1)}
=
\Pi_{[C_{\min},C_{\max}]}
\left\{
C^{(t)}
-\clip\!\left[
\frac{g_C(A,C^{(t)})}{F_{CC}(A,C^{(t)})},
-\Delta_C,\Delta_C
\right]
\right\},
\end{equation}
where \(\Pi\) denotes projection onto the stated interval{,
$\clip[x,a,b]=\min\{\max(x,a),b\}$,} and \(\Delta_C\)
limits the magnitude of one update. The inner iteration terminates when
$|C^{(t+1)}-C^{(t)}|\le\tau\,(1+|C^{(t)}|)$ or after $N_{\mathrm{inner}}$
updates.

\subsubsection{Outer optimization}

For an interior and differentiable inner solution, the envelope theorem gives
\begin{equation}
\frac{\mathrm{d}\mathcal{L}_{\mathrm{p}}(A)}{\mathrm{d}A}
=g_A\!\left[A,C^{\star}(A)\right].
\label{eq:profile_gradient}
\end{equation}
{As in Fisher scoring, the curvature of the profile objective is approximated by the Schur complement of the Fisher information matrix,}
\begin{equation}
H_{\mathrm{p}}
=F_{AA}-\frac{F_{AC}^{2}}{F_{CC}}.
\label{eq:profile_curvature}
\end{equation}
The outer update is
\begin{equation}
\label{eq:outer-update}
A^{(s+1)}
=
\Pi_{[A_{\min},A_{\max}]}
\left\{
A^{(s)}
-\clip\!\left[
\frac{g_A(A^{(s)},C_s)}
{H_{\mathrm p}(A^{(s)},C_s)},
-\Delta_A,\Delta_A
\right]
\right\}.
\end{equation}
Here \(C_s\) is the result of the finite inner iteration{. No} claim is made that the
global inner minimizer has been obtained exactly. The outer iteration uses the
same stopping test on \(A\), with limit \(N_{\mathrm{outer}}\). The inner coefficient is
reprofiled after the final \(A\) update so that the returned pair corresponds
to the final outer value. Curvature floors ($F_{CC}$ and $H_{\mathrm p}$ at $10^{-12}$), parameter
bounds, step limits, and finite iteration ceilings are used as numerical
safeguards. A ray is counted as in contact with a bound when an estimate lies
within $2\times10^{-4}$~g\,cm$^{-2}$ of it. Non-finite
or poorly conditioned final states are reported through the solver
diagnostics.

The numerical controls used by the accelerated implementation are
\begin{equation}
\label{eq:numerical-controls}
\begin{aligned}
[A_{\min},A_{\max}]&=[-0.10,0.40]~\mathrm{g\,cm^{-2}},
&[C_{\min},C_{\max}]&=[-2.0,50.0]~\mathrm{g\,cm^{-2}},\\
\Delta_A&=0.05~\mathrm{g\,cm^{-2}},
&\Delta_C&=5.0~\mathrm{g\,cm^{-2}},\\
\tau&=5\times10^{-5},
&(N_{\mathrm{outer}},N_{\mathrm{inner}})&=(16,12).
\end{aligned}
\end{equation}

The total-count refinement of Eq.~\eqref{eq:bisection_initializer} uses 28 bisection iterations. The small negative
lower bounds reduce truncation of noisy estimates near zero and are not a
physical assertion of negative material density. The bounds, step limits,
and iteration ceilings are numerical controls rather than spatial
regularization or denoising parameters.

Each ray returns the estimated basis pair, the Fisher elements, inner and
outer iteration counts, and indicators for convergence, boundary contact,
aggregate feasibility, conditioning, and non-finite model states. The
normalized final gradient norm is
\begin{equation}
\label{eq:normalized-score}
\kappa_g
=
\frac{\sqrt{g_A^2+g_C^2}}
{\sqrt{\max(F_{AA}+F_{CC},10^{-12})}}.
\end{equation}
The implementation flags the Fisher matrix as poorly conditioned when its
eigenvalue condition number exceeds \(10^8\). Both quantities depend on the
fixed scaling of the two basis functions and are numerical diagnostics, not
basis-invariant hypothesis tests.

An optional air gate sets both basis line integrals to zero, bypassing the
optimization, for any ray whose largest absolute log measurement falls below
a threshold. The threshold is 0 for all reported results except its sweep in
{Section~\ref{sec:results_ablation}}.

\subsection{Bounded numerical reference}
\label{sec:reference_solver}

The accelerated updates are not used as their own correctness standard. A
bounded reference solver samples 65 values of $C$ over
$[C_{\min},C_{\max}]$ at every tested $A$, refines each detected basin by 80
golden-section iterations, evaluates both endpoints, and treats the sampled
outer profile over $[A_{\min},A_{\max}]$ in the same way. This reference is
independent of the initialization and update sequence of
Section~\ref{sec:nested_optimization}{. Agreement} with it is reported in the
evaluation.

\subsection{Denoising and VMI generation}
\label{sec:denoising_vmi}

For the photon-counting acquisition, the estimated basis sinograms were
first denoised using the total-likelihood bilateral
filter~\cite{LeeKim2025}, an extension of the likelihood-based bilateral
filter of Lee~\cite{Lee2023}. The filter used a $5\times5$ neighborhood,
a spatial parameter $\alpha_{1}=0.9$ detector pixels, and a likelihood
parameter $\alpha_{2}=24.6$, in the notation of Lee and Kim~\cite{LeeKim2025}.
Circular wrapping was applied along the projection dimension, whereas the
detector boundaries were not wrapped. Because the filter weights neighbors
by the Poisson likelihood of the measured photon counts, which
energy-integrating detectors do not record, it was not applied to the
energy-integrating acquisitions.

For every acquisition, the basis sinograms were then reconstructed
independently using filtered backprojection (FBP), and anticorrelated noise
reduction was applied jointly to the two basis images~\cite{Kalender1988}.

VMIs were then generated from the denoised
basis images at energy $E_0$ as
\begin{equation}
\mu_{\mathrm{VMI}}(\boldsymbol{r},E_0)
=
f_1(E_0)\widetilde{a}(\boldsymbol{r})
+
f_2(E_0)\widetilde{c}(\boldsymbol{r}),
\label{eq:vmi_generation}
\end{equation}
where $\widetilde{a}(\boldsymbol{r})$ and
$\widetilde{c}(\boldsymbol{r})$ denote the denoised iodine and water density
maps (g\,cm$^{-3}$). CT numbers{, in Hounsfield units (HU),} were computed as
$\mathrm{HU}=1000\,[\mu_{\mathrm{VMI}}(\boldsymbol r,E_0)-\mu_{\mathrm w}(E_0)]/\mu_{\mathrm w}(E_0)$,
with $\mu_{\mathrm w}(E_0)=f_2(E_0)\times1~\mathrm{g\,cm^{-3}}${. The} calculated
reference values of Section~\ref{sec:eval_metrics} use the same definition.

\subsection{Experiments}
\label{sec:experiments}

The proposed {method} was evaluated at three levels. First, the accuracy
and statistical efficiency of the material decomposition were assessed
using simulated data for which the true water and iodine line
integrals were known. The recovered line integrals were compared with the
ground truth and with a bounded numerical reference, and their variability
over independent noise realizations was compared with the CRLB. Second, quantitative accuracy was evaluated on VMIs, with reference CT numbers calculated from the
known material compositions and tabulated attenuation coefficients. Third,
the simulated VMI HU values were compared with those measured from physical
scans of the same phantom.

VMIs serve here as the readout of the decomposition rather than as its
product. A VMI is a fixed linear combination of the two basis images
(Eq.~\eqref{eq:vmi_generation}), so its CT numbers inherit any error in the
estimated line integrals. VMIs are also the form in which the physical
systems report their spectral results, which makes them the quantity that can
be compared with a physical scan.

The direct decomposition analysis was performed on the estimated water and
iodine sinograms before filtering or reconstruction. VMI accuracy and noise
were evaluated after the processing described in
Section~\ref{sec:denoising_vmi}. Within each acquisition, this processing was
applied identically to all decomposition methods and was not considered part
of the proposed decomposition {method}. Physical scan VMIs{, generated on each scanner by the manufacturer's own
software,} were used for
comparison of CT-number accuracy. Their noise is shown for reference only, because
the manufacturer-specific decomposition and noise-reduction processing are not
available and therefore do not permit a controlled comparison of image noise.

\begin{figure}[!htbp]
\centering
\includegraphics[width=\textwidth]{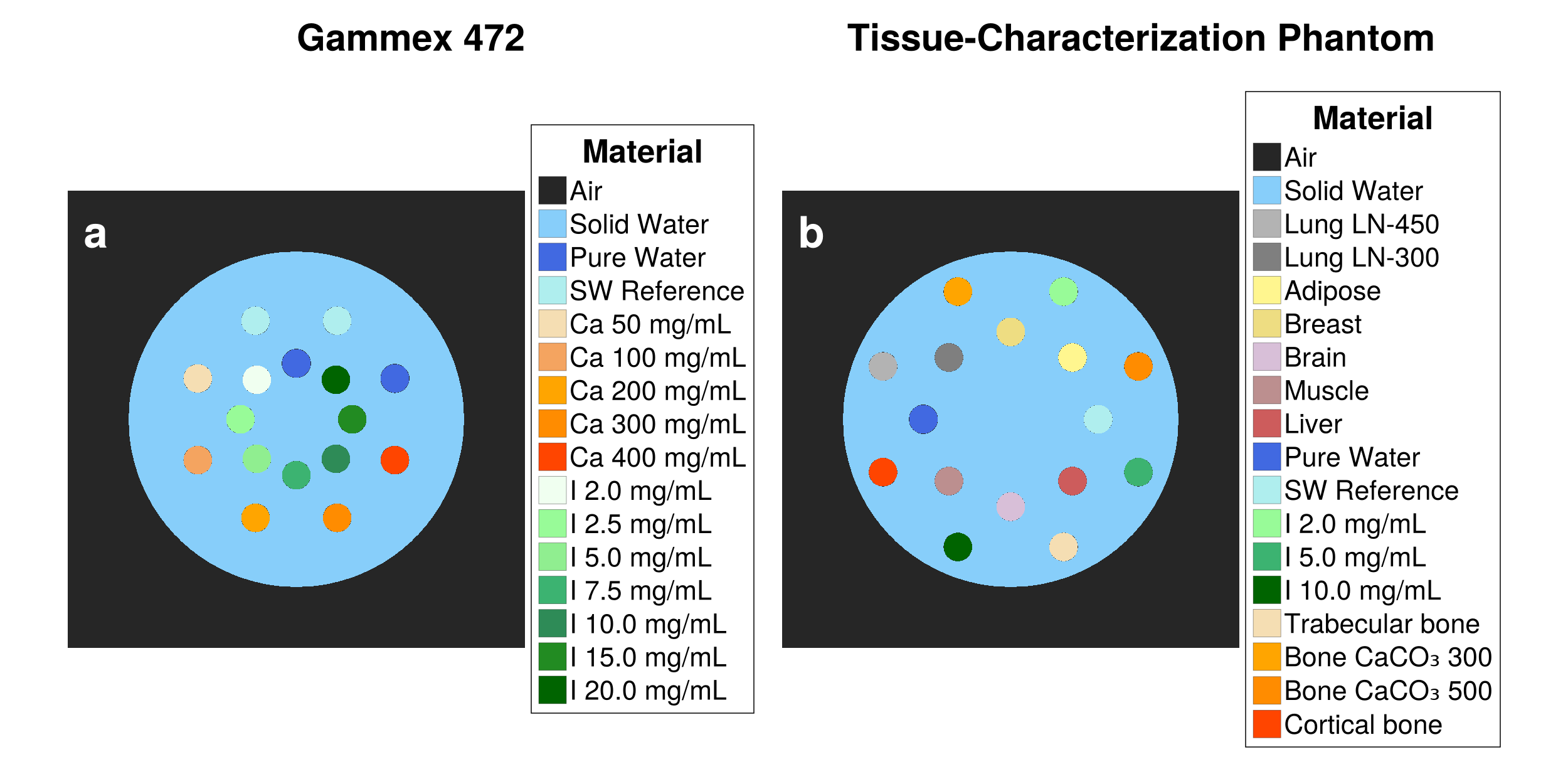}
\caption{Simulated phantoms used for evaluation.
(a) Gammex 472 phantom, {in} the configuration used for the physical
scans, with pure-water, solid-water, iodine (2--20~mg/mL), and calcium
(50--400~mg/mL) inserts.
(b) Tissue-characterization phantom containing materials spanning
approximately $-805$ to $1559$~HU at 70~keV. The Gammex 472 phantom was
evaluated using both simulated and physical acquisitions, whereas the
tissue-characterization phantom was evaluated in simulation only.}
\label{fig:phantoms}
\end{figure}

\subsubsection{Simulated acquisitions and phantoms}
\label{sec:eval_setup}

All simulations were performed using \texttt{BasisSimulator.jl} version 0.14.0~\cite{Black2026}. Three spectral CT acquisition configurations, each based on a different
spectral acquisition mechanism~\cite{McCollough2015}, were evaluated: two energy-integrating systems, the Revolution Apex Elite
(GE HealthCare, Waukesha, WI, USA) with rapid kVp switching and the SOMATOM
Definition Flash dual-source CT system (Siemens Healthineers, Forchheim,
Germany), and the NAEOTOM Alpha photon-counting CT system (Siemens
Healthineers). The GE acquisition alternated between 80 and
140~kVp. The SOMATOM Flash used 80 and
140~kVp spectra, with 0.4~mm Sn filtration applied to the high-energy
spectrum. The NAEOTOM Alpha was modeled at 140~kVp using four energy windows
with thresholds of 20, 35, 55, and 70~keV. For all three systems, the energy-dependent detector response was obtained
from Monte Carlo radiation-transport simulations of the corresponding
detector~\cite{Khodajou2019}. For the Gemstone scintillator and the {UltraFast Ceramic (UFC)
gadolinium oxysulfide (Gd$_2$O$_2$S)} scintillator of the two
energy-integrating systems, these simulations give the absorption efficiency
at each energy, including fluorescence escape at the K-edges of the
scintillator elements. For the CdTe detector of the NAEOTOM Alpha, they give
the full energy-response matrix, including charge sharing and fluorescence
escape. The corresponding acquisition and reconstruction parameters are
summarized in Table~\ref{tab:acquisitions}.

Two phantoms were used in the simulation study: the Gammex 472 phantom (Gammex Inc., Middleton, WI, USA)~\cite{Gammex472} and a tissue-characterization phantom, shown in
Fig.~\ref{fig:phantoms}. The Gammex 472 phantom was also scanned physically
on all three systems. The simulated acquisitions used the tube potential, tube current, and rotation
time of the corresponding physical scans. {The decomposition used the
spectral responses that generated the simulated data, except where
Section~\ref{sec:eval_pcct} states otherwise. The two measurements of each
dual-energy acquisition were simulated along the same ray paths.} Because the material compositions of the phantom inserts were known,
reference monoenergetic attenuation values were calculated using NIST XCOM
mass-attenuation coefficients~\cite{NISTXCOM}.

\begin{table}[!htbp]
\centering
\caption{Acquisition parameters used for the simulated spectral CT
configurations. Tube potential, tube current, and rotation time {are those of
the} corresponding physical Gammex 472 acquisitions, and CTDI$_{\mathrm{vol}}$ {(volume CT
dose index)} is the value reported for those acquisitions. {The field of
view is that of the simulated reconstruction. The SOMATOM Flash physical scan
was reconstructed over its 33~cm dual-energy field of view.}}
\label{tab:acquisitions}
\small
\begin{tabular}{lccc}
\toprule
 & \multicolumn{2}{c}{Energy-Integrating} & Photon-Counting\\
\cmidrule(lr){2-3}\cmidrule(lr){4-4}
 & GE Revolution & Siemens SOMATOM & Siemens NAEOTOM\\
Parameter & Apex Elite & Flash & Alpha\\
 & (dual-kVp, $K=2$) & (dual-source, $K=2$) & (PCCT, $K=4$)\\
\midrule
Spectral mechanism       & rapid kVp    & dual          & photon\\
                         & switching    & source        & counting\\
Detector material        & Gemstone     & {Gd$_2$O$_2$S (UFC)} & CdTe\\
Detector thickness (mm)  & 3.0          & {1.0}           & 1.6\\
Source-to-isocenter (mm) & 625.6        & 595.0         & 610.0\\
Source-to-detector (mm)  & 1100.0       & 1085.6        & 1113.0\\
Detector pitch at isocenter (mm) & $0.600\times0.625$ & $0.705\times0.600$ & $0.301\times0.353$\\
Tube potential (kVp)     & 80 / 140     & 80 / Sn140    & 140\\
Tube current, instantaneous (mA) & 407 / 405 & 470 / 182 & 174\\
Duty cycle               & 0.65 / 0.35  & 1.0 / 1.0     & 1.0\\
Gantry rotation time (s) & 0.5          & 0.5           & 0.5\\
Views per rotation       & 984 per kVp  & 1152 per tube & 1200\\
Permanent filtration     & Al 2.5\,mm   & Al 8.4\,mm    & Al 3.0\,mm\\
Added filtration         & Al 4.5\,mm   & Sn 0.4\,mm (tube B) & Ti 0.9\,mm\\
Energy thresholds (keV)  & ---          & ---           & 20 / 35 / 55 / 70\\
CTDI$_{\mathrm{vol}}$, physical scan (mGy) & 10.07 & 10.01 & 10.12\\
Reconstructed matrix     & $512^2$      & $512^2$       & $512^2$\\
Field of view (cm)       & 35           & 35            & 35\\
\midrule
Physical scan of same phantom & yes     & yes           & yes\\
\bottomrule
\end{tabular}
\end{table}

\subsubsection{Benchmarking}
\label{sec:eval_benchmark}

The proposed {method} was benchmarked against the univariate
projection-domain decomposition of Cong et al.~\cite{Cong2022}, the two-measurement
method that it generalizes (Section~\ref{sec:nested_optimization}).
For the two-measurement dual-energy acquisitions, both methods were applied
directly to the same spectral measurements. For the photon-counting
acquisition, the proposed method was applied {both to all four energy windows
($K=4$) and to} two measurements formed by {merging} windows 1--2
and 3--4 {($K=2$)}. Each {merged} measurement is the sum of the counts, and of the spectral
responses, of its windows. {The univariate method used the same two merged
measurements, so comparing the three} separates the effect of the number of
measurements from that of the algorithm.

The univariate method was applied using its implementation in
\texttt{BasisSimulator.jl}~\cite{Black2026}, which follows the original
method~\cite{Cong2022}, with Newton's method for the polynomial root and
Brent's method for the one-dimensional search. Both methods used the same input data and
spectral models and were followed by identical reconstruction and
image-processing steps.

\subsubsection{Photon-counting corrections}
\label{sec:eval_pcct}

\begin{figure}[!htbp]
\centering
\includegraphics[width=0.6\textwidth]{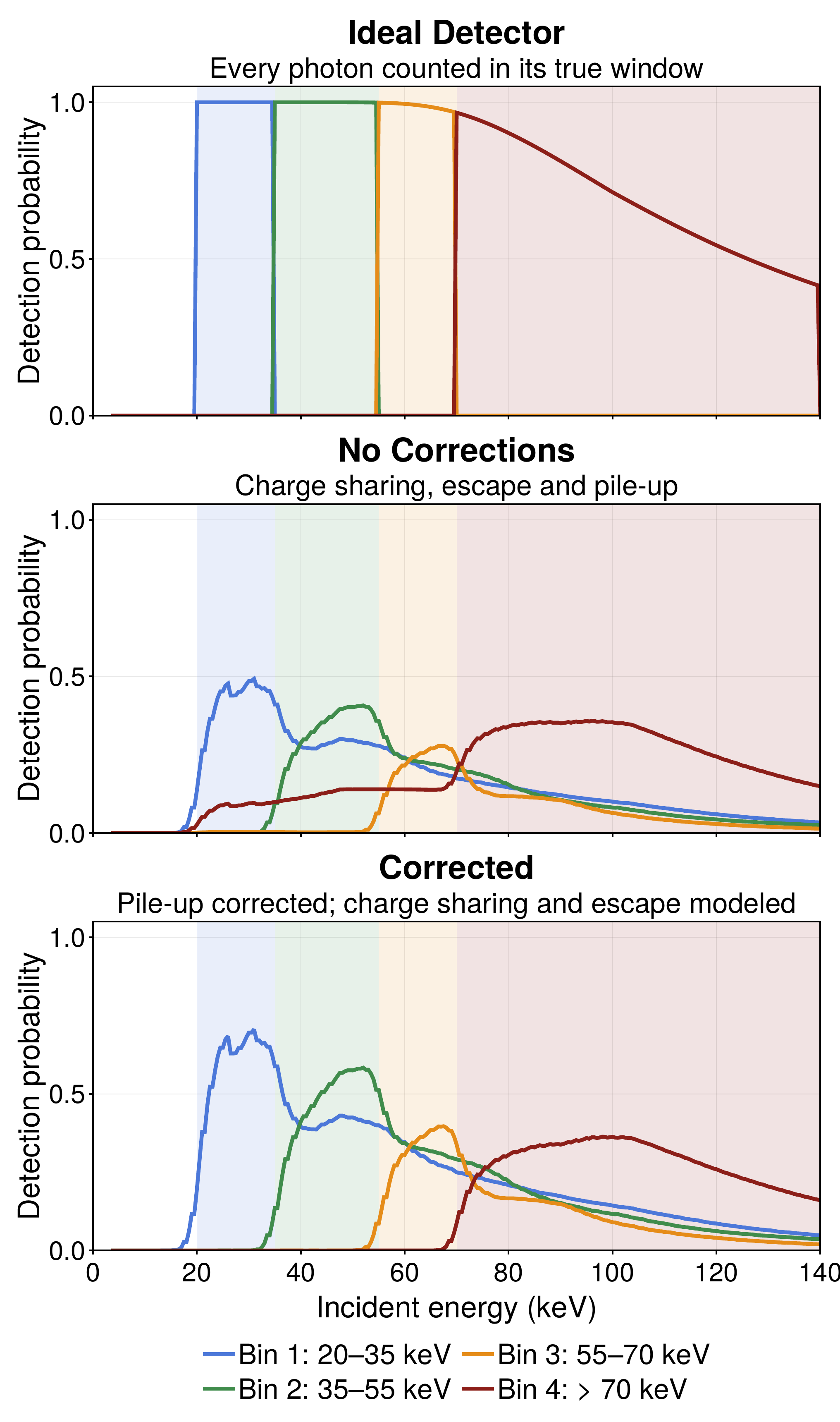}
\caption{Detector response of the four energy windows of the photon-counting
acquisition: the probability that a photon of a given incident energy is
counted in each window. Top: an ideal detector, in which every photon is
counted in the window of its true energy. Middle: the uncorrected detector,
with charge sharing, fluorescence escape and pile-up. Bottom: after pile-up
correction, where the charge-sharing and escape tails remain and are modeled
by the decomposition. Shaded bands mark the nominal windows.}
\label{fig:pcct_corrections}
\end{figure}

The decomposition models the detector only through the spectral response
$S_k(E)$ (Eq.~\eqref{eq:continuous_forward}), so its accuracy depends on how
well that response describes the measured counts. The photon-counting
simulation models many physical effects, and this dependence was examined for
four of them that change the counts recorded in each window. Charge sharing and fluorescence escape lower the energy at
which a photon is recorded, and pulse pile-up merges near-coincident photons
into one event of higher apparent energy. All three move counts between
energy windows. Patient scatter instead adds photons that depend on the
object. In the reference configuration, pile-up was corrected by inverting
the Monte Carlo pile-up model, scatter was corrected by re-estimating the
scatter field from the measured windows, and charge sharing and escape, which
cannot be removed from the counts, were included in the detector response in
$S_k(E)$. Figure~\ref{fig:pcct_corrections} shows the detector response of each
window, the probability that a photon of a given incident energy is counted in
it, for an ideal detector, for the uncorrected detector, and after pile-up
correction.

To show what the decomposition depends on, the Gammex 472 acquisition was
repeated with no corrections: pile-up and scatter were left in the counts,
and its line integrals were referenced to an air scan that was itself piled
up, as on a scanner without pile-up correction. Because the counts are drawn
before pile-up and both corrections, this acquisition, with the same seed,
has the same noise realization as the reference. It was decomposed twice:
with an ideal response, in which every photon is counted in the window of its
true energy, and with the response adjusted to include pile-up. In the simulator, pile-up
is a fixed migration of counts between windows, with some counts lost, so it
can be folded into the expected counts through $S_k(E)$. Scatter cannot be represented by any response and remained in
both. The corrections and the adjusted response use the same pile-up,
scatter, and detector models that generated the data, so they represent the
best case. With measured data, each would be a calibrated estimate. Each configuration was decomposed with the proposed method using the
four windows and using the same two merged windows as above ($K=2$), with
otherwise identical processing.

\subsubsection{Evaluation metrics}
\label{sec:eval_metrics}

Performance was evaluated in terms of decomposition accuracy, VMI accuracy
against reference CT numbers, image noise, statistical efficiency, numerical
sensitivity, and decomposition time.

The decomposition was first verified on noiseless data. Measurements were
generated from 145 known $(A,C)$ pairs covering the path lengths in the
phantoms: water alone with $0\le C\le35$~g\,cm$^{-2}$, and iodine with
$0.005\le A\le0.06$~g\,cm$^{-2}$ and $10\le C\le35$~g\,cm$^{-2}$. The
maximum iodine error, $\max|\widehat A-A|$, is reported over all pairs. For 21
of the pairs, the final objective value was also compared with the minimum
found by the bounded reference of Section~\ref{sec:reference_solver}, and the
relative difference is reported.

VMI accuracy was quantified for each insert as the difference between the
mean HU within the insert {region of interest (ROI)} of the simulated VMI and a reference value,
\begin{equation}
\Delta\mathrm{HU}_{\mathrm{ref}}(E_0)
=
\overline{\mathrm{HU}}_{\mathrm{sim}}(E_0)
-
\mathrm{HU}_{\mathrm{ref}}(E_0),
\qquad
\mathrm{ref}\in\{\mathrm{Calc.},\mathrm{Phys.}\},
\label{eq:hu_error}
\end{equation}
where $\mathrm{HU}_{\mathrm{Calc.}}$ is calculated from the insert's known
composition with NIST XCOM coefficients~\cite{NISTXCOM} and
$\mathrm{HU}_{\mathrm{Phys.}}$ is the mean HU in the same ROI of the physical
scan VMI. Accuracy was summarized at each energy over the inserts by the mean
absolute error (MAE), the mean of $|\Delta\mathrm{HU}_{\mathrm{ref}}|$, and by
the maximum absolute error (Max.\ error). Simulated HU values
were also regressed on the reference values by ordinary least squares,
separately at each VMI energy. The slope and the root-mean-square residual
about the fitted line (the residual error) are reported.

Image noise was measured on the synthesized VMIs as the standard deviation of
HU, $\sigma_{\mathrm{HU}}$, within a uniform solid-water region and as the
mean within-insert standard deviation (SD) over the inserts of each phantom.

Statistical efficiency was evaluated on the estimated line integrals, before
denoising or reconstruction, for a representative ray through the phantom
body with $(A,C)=(0.02,25)$~g\,cm$^{-2}$. For each acquisition and method,
20{,}000 independent Poisson realizations of the expected counts were
generated, and counts were floored at 0.5 before the logarithm. The empirical
standard deviation was then normalized by the corresponding CRLB {standard deviation},
\begin{equation}
\eta_A=
\frac{\widehat{\sigma}_A}
{\sqrt{[\boldsymbol F^{-1}]_{AA}}},
\qquad
\eta_C=
\frac{\widehat{\sigma}_C}
{\sqrt{[\boldsymbol F^{-1}]_{CC}}}.
\label{eq:efficiency}
\end{equation}
Values near unity indicate that the empirical variability approaches the
{lower bound} predicted by the measurement model. Because the CRLB applies only to
unbiased estimators, the normalized bias (the mean error divided by
$\widehat\sigma_A$) and the percentage of estimates at a parameter bound are
reported alongside $\eta$. Beyond sampling uncertainty, a value of $\eta$
below unity therefore indicates bias{, for example from bound contact or the count floor,} not an
estimator that outperforms the bound.

Numerical sensitivity was assessed on the NAEOTOM Alpha Gammex 472
acquisition. Each control was set to each end of its range while the others
kept their default values. In {Section~\ref{sec:results_ablation}}, $\Delta\mathrm{HU}$ is
the change in insert HU from the default settings, not an error against a
reference.

\begin{figure}[!htbp]
\centering
\includegraphics[width=\textwidth]{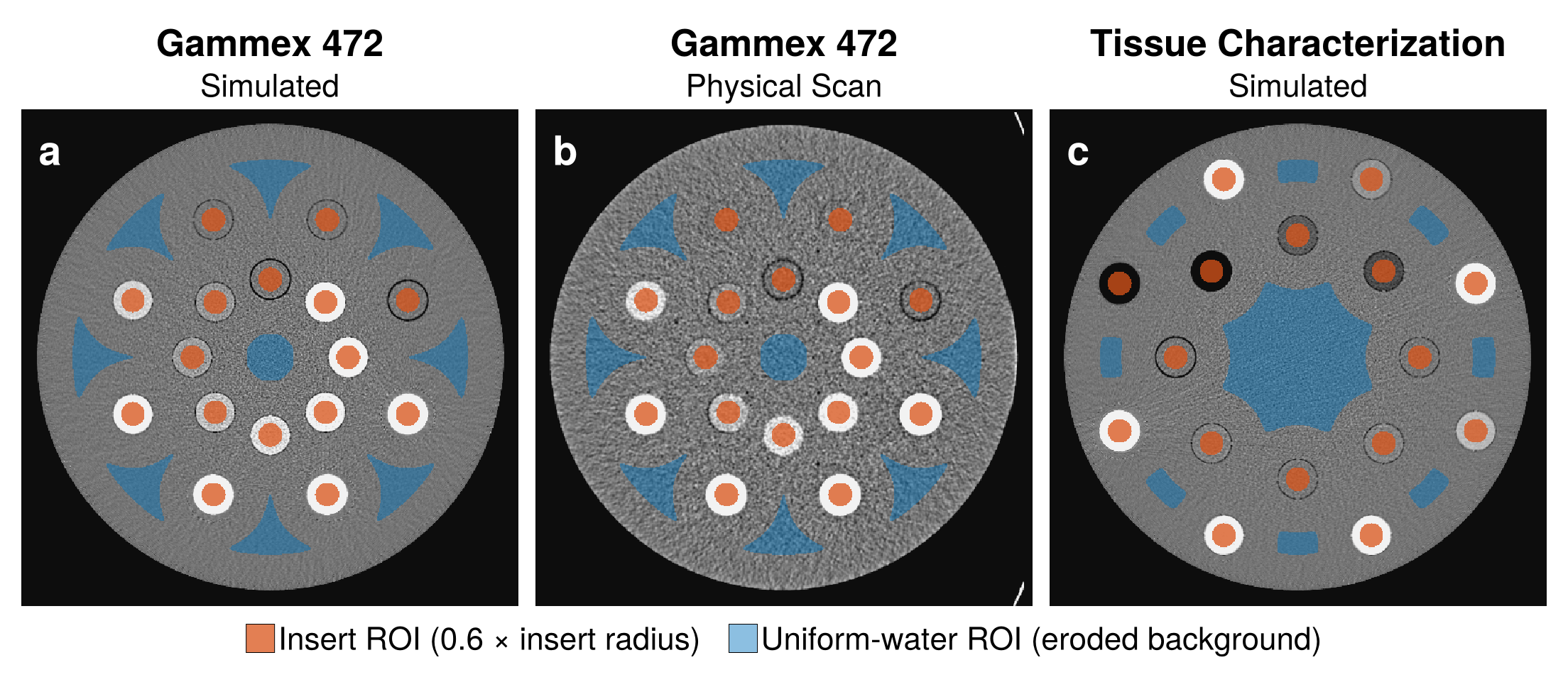}
\caption{Regions of interest used for the CT-number and noise measurements,
shown on 70~keV virtual monoenergetic images. (a) Simulated Gammex 472 phantom (NAEOTOM Alpha,
$K=4$). (b) Physical NAEOTOM Alpha scan of the same phantom, onto which the
regions were transferred after registration. (c) Simulated
tissue-characterization phantom. Insert regions (orange) have 0.6 times the
insert radius{. The} uniform-water region (blue) is the solid-water background
after an 8~mm margin is excluded from the phantom boundary and each insert.
Display window $[-200,240]$~HU.}
\label{fig:rois}
\end{figure}

Insert measurements were obtained from circular {ROIs} with a radius of 8.4~mm
(0.6 times the insert radius), corresponding to approximately
470 pixels for the 0.68~mm reconstructed pixel size (0.645~mm for the SOMATOM
Definition Flash physical scan{, reconstructed over its 33~cm dual-energy
field of view}). {The insert regions lie well inside it.} The uniform-water ROI
was defined within the solid-water background after excluding an 8~mm
margin from the phantom boundary and each insert. The same ROIs were
transferred to the physical scan images after each scan was registered to the
phantom model in position, orientation, and rotation (Fig.~\ref{fig:rois}).

Simulated measurements were obtained from a single noise realization of the
reconstructed slice. Physical scan measurements were obtained from the
central slice of each vendor VMI series: 0.625~mm with the standard kernel for
the GE Revolution Apex Elite, 1.0~mm with the D40f kernel for the SOMATOM
Definition Flash, and 0.4~mm with the Br36f kernel for the NAEOTOM Alpha.

Decomposition time was measured on an NVIDIA A100 GPU, excluding
reconstruction and image processing, and is reported per million rays as the
median of repeated runs. For comparison with the bounded reference, both
solvers were also timed on the CPU over the same 120 rays of the NAEOTOM Alpha
sinogram.

\section{Results}
\label{sec:results}

\subsection{Decomposition performance}
\label{sec:results_solver}

Table~\ref{tab:solver} summarizes the evaluation of the proposed {method} in the sinogram domain. For noiseless measurements, the
proposed method recovered iodine line integrals to within
$2.7\times10^{-6}$~g\,cm$^{-2}$ across all evaluated configurations. Its
final objective values agreed with those obtained using the bounded
numerical reference to a relative difference of $3.4\times10^{-13}$ or less,
with no non-finite or bound-limited solutions. The corresponding worst cases
for the univariate method were $1.6\times10^{-5}$~g\,cm$^{-2}$ and
$2.3\times10^{-11}$. These results verify that
the nested optimization converges to the expected solution within the
selected numerical tolerances.

Statistical performance was evaluated over 20{,}000 independent Poisson
realizations. The empirical variability of the recovered water and iodine
line integrals remained within 6\% of the corresponding CRLB across the
six evaluated acquisition--method combinations, with an uncertainty in the normalized
standard deviation of approximately $\pm0.5\%$. The estimates showed
{a normalized bias of at most 0.11 and no appreciable} interaction with the parameter bounds.

For the photon-counting acquisition, the effect of retaining the additional
spectral measurements was also evident from the CRLB. Combining the
four energy windows into two measurements increased the iodine CRLB standard
deviation from 0.0446 to 0.0549~g\,cm$^{-2}$ for the representative ray,
corresponding to an increase of approximately 23\% {(a 19\% reduction with
$K=4$)}. Each evaluated estimator
approached its own bound ($\eta_A=1.03$ for the proposed method with $K=4$ and
1.06 for the univariate method with $K=2$), indicating that the
difference arises primarily from the information lost when the four
measurements are reduced to two.

The computational cost remained low across the evaluated configurations.
{The proposed method required 0.34--0.67~s per million rays on the GPU.} The bounded numerical
reference required approximately 60 times the computation time when both
methods were evaluated on the CPU. A complete reference calculation was
estimated to require approximately 47~h based on timing a representative
subset of 120 rays.

\begin{table}[!htbp]
\centering
\caption{Quantitative evaluation of the proposed {method} in the sinogram domain.
The noiseless evaluation reports the maximum error in the recovered iodine
line integral and the relative objective-function difference from the
bounded numerical reference. Statistical performance over 20{,}000
independent Poisson realizations is characterized by the iodine {CRLB
standard deviation}, CRLB-normalized standard deviations for iodine and water, normalized iodine
bias, and the fraction of estimates reaching a parameter bound.
The univariate method brackets its searches rather than bounding the
parameters, so bound contact does not apply to it (---). Decomposition time
is reported per million rays on an NVIDIA A100 GPU.}
\label{tab:solver}

\footnotesize
\setlength{\tabcolsep}{2.5pt}

\begin{tabular}{llrrrrrrrr}
\toprule
& &
\multicolumn{2}{c}{Noiseless} &
\multicolumn{5}{c}{20{,}000 Poisson realizations} &
\\
\cmidrule(lr){3-4}
\cmidrule(lr){5-9}

Acquisition & Method
& $\max|\widehat A-A|$
& Obj.\ gap
& CRLB$_A$
& $\eta_A$
& $\eta_C$
& Bias$_A$
& At bound
& Time \\

& &
(g\,cm$^{-2}$)
& &
(g\,cm$^{-2}$)
& &
&
($\widehat\sigma_A$)
& (\%)
& (s/Mray) \\

\midrule

\multicolumn{10}{l}{\textit{Energy-Integrating}}\\

GE Revolution ($K=2$)
& Proposed
& $1.5\times10^{-6}$
& $1.9\times10^{-13}$
& 0.0207
& 1.01
& 1.00
& 0.03
& 0.0
& 0.55 \\

GE Revolution ($K=2$)
& Univariate
& $1.6\times10^{-5}$
& $1.4\times10^{-11}$
& 0.0207
& 1.01
& 1.00
& 0.03
& ---
& 0.31 \\

SOMATOM Flash ($K=2$)
& Proposed
& $5.7\times10^{-7}$
& $7.1\times10^{-14}$
& 0.0151
& 1.00
& 1.00
& 0.01
& 0.0
& 0.67 \\

SOMATOM Flash ($K=2$)
& Univariate
& $8.3\times10^{-6}$
& $2.3\times10^{-11}$
& 0.0151
& 1.00
& 1.00
& 0.02
& ---
& 0.30 \\

\midrule
\multicolumn{10}{l}{\textit{Photon-Counting}}\\

NAEOTOM Alpha ($K=4$)
& Proposed
& $2.7\times10^{-6}$
& $3.4\times10^{-13}$
& 0.0446
& 1.03
& 1.02
& 0.04
& 0.1
& 0.34 \\

NAEOTOM Alpha ($K=2$)
& Univariate
& $2.6\times10^{-6}$
& $5.3\times10^{-13}$
& 0.0549
& 1.06
& 1.02
& 0.11
& ---
& 0.18 \\

\bottomrule
\end{tabular}
\end{table}%

\subsection{VMI evaluation}
\label{sec:results_theory}

\begin{figure}[!htbp]
\centering
\includegraphics[width=\textwidth]{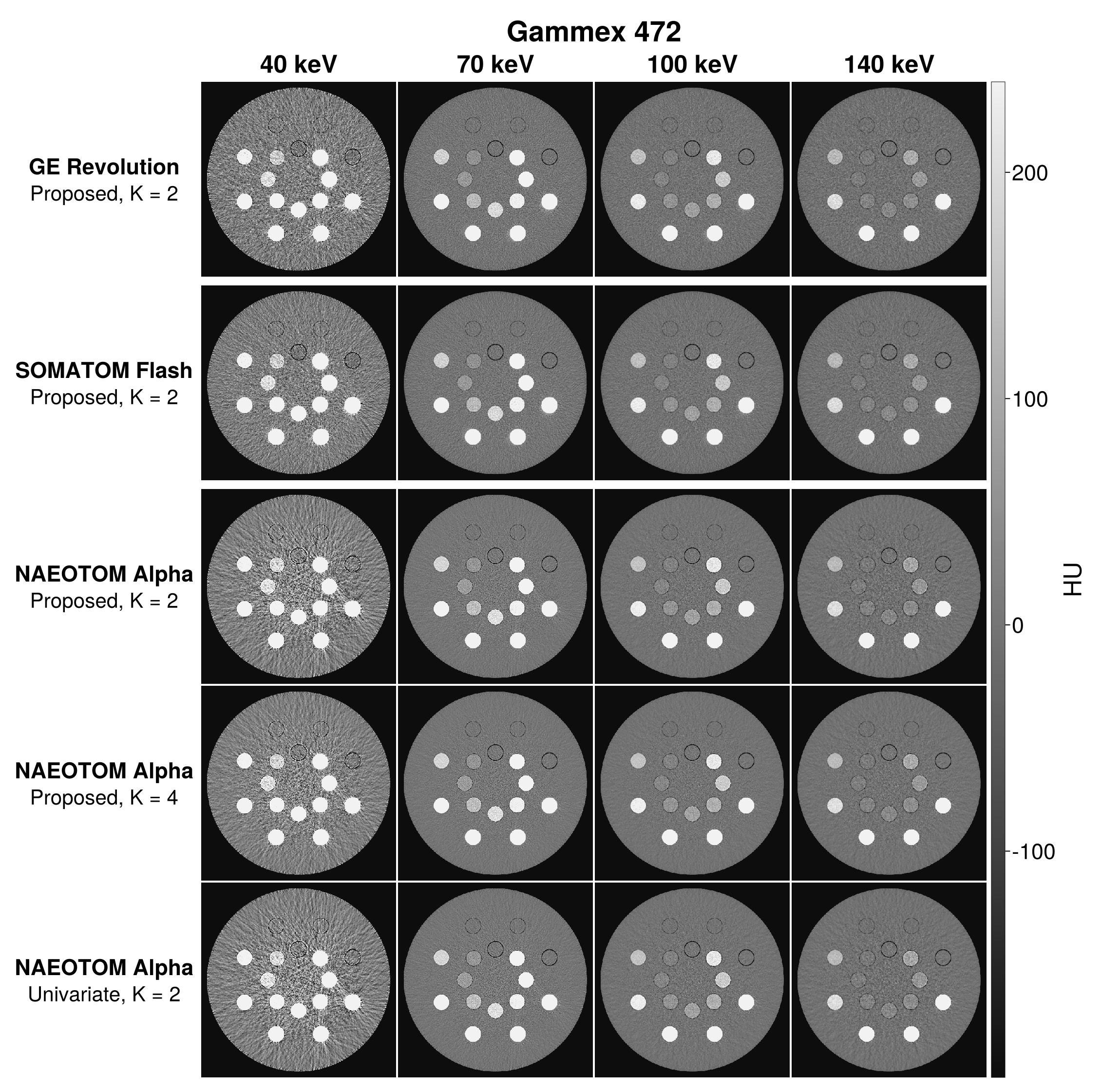}
\caption{Virtual monoenergetic images of the Gammex 472 phantom at
40, 70, 100, and 140~keV. Rows show the proposed method for the
GE Revolution Apex Elite ($K=2$), Siemens SOMATOM Flash ($K=2$),
and Siemens NAEOTOM Alpha ($K=2$ and $K=4$), followed by the
univariate method for the NAEOTOM Alpha ($K=2$). All images are
displayed using the same $[-200,240]$~HU window.}
\label{fig:vmi_gammex}
\end{figure}

\begin{figure}[!htbp]
\centering
\includegraphics[width=\textwidth]{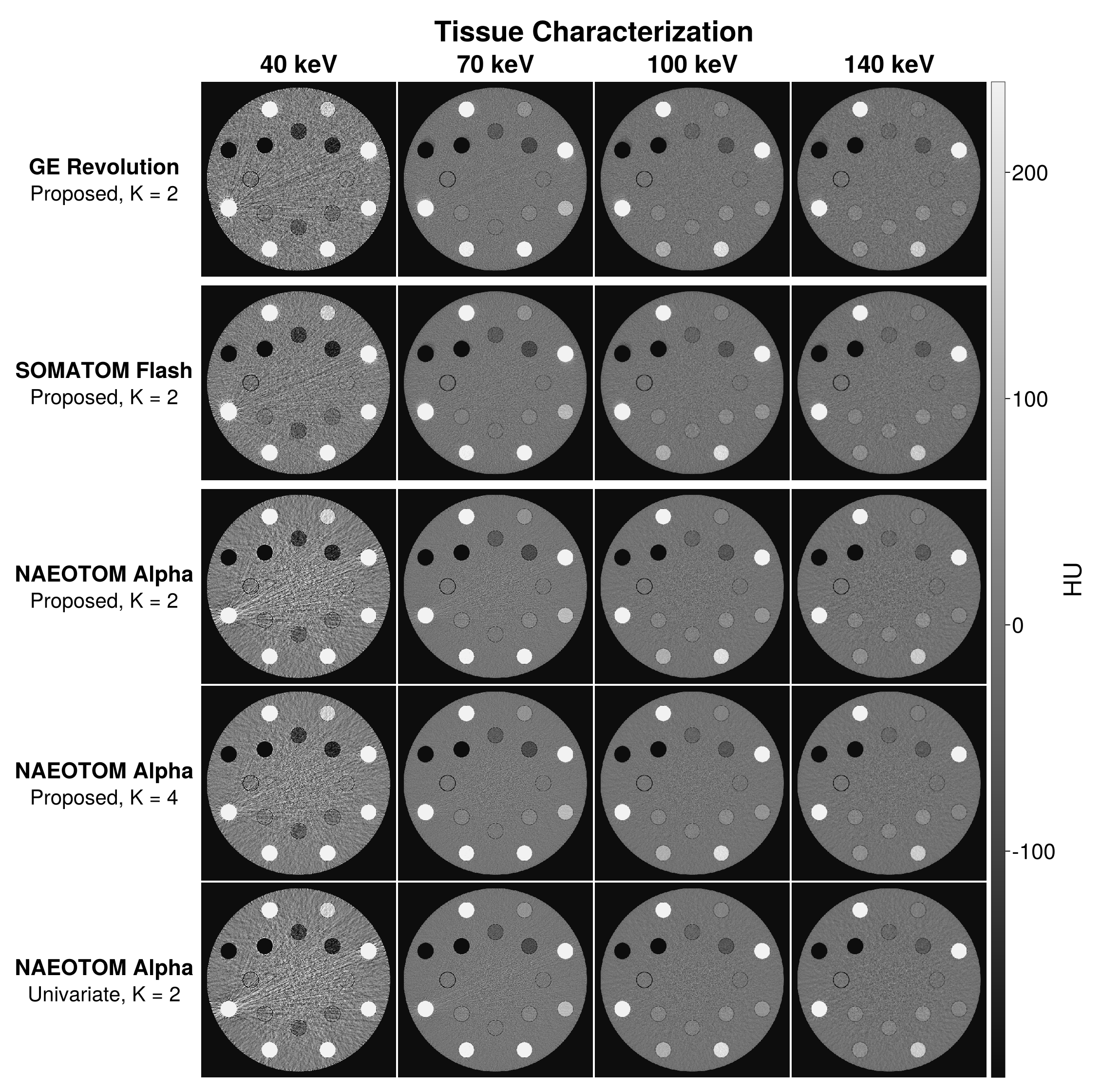}
\caption{Virtual monoenergetic images of the simulated
tissue-characterization phantom at 40, 70, 100, and 140~keV. Rows show
the proposed method for the GE Revolution Apex Elite ($K=2$), Siemens
SOMATOM Flash ($K=2$), and Siemens NAEOTOM Alpha ($K=2$ and $K=4$),
followed by the univariate method for the NAEOTOM Alpha ($K=2$).
All images are displayed using the same $[-200,240]$~HU window as
Fig.~\ref{fig:vmi_gammex}.}
\label{fig:vmi_tissue}
\end{figure}

Figures~\ref{fig:vmi_gammex} and~\ref{fig:vmi_tissue} show the reconstructed
VMIs for the Gammex 472 and tissue-characterization phantoms, respectively.
The proposed method produced consistent material contrast across the three
acquisition configurations. Noise and streak artifacts were most visible
at the lowest VMI energy. For the photon-counting acquisition, using all
four spectral measurements reduced the visible noise relative to the
two-measurement implementation, with the difference being most apparent at
40~keV. A reduction in streaking was also visible around the high-density
bone inserts in the tissue-characterization phantom. The proposed method ($K=2$) and
the univariate method produced visually similar images.

\begin{figure}[!htbp]
\centering
\includegraphics[width=\textwidth]{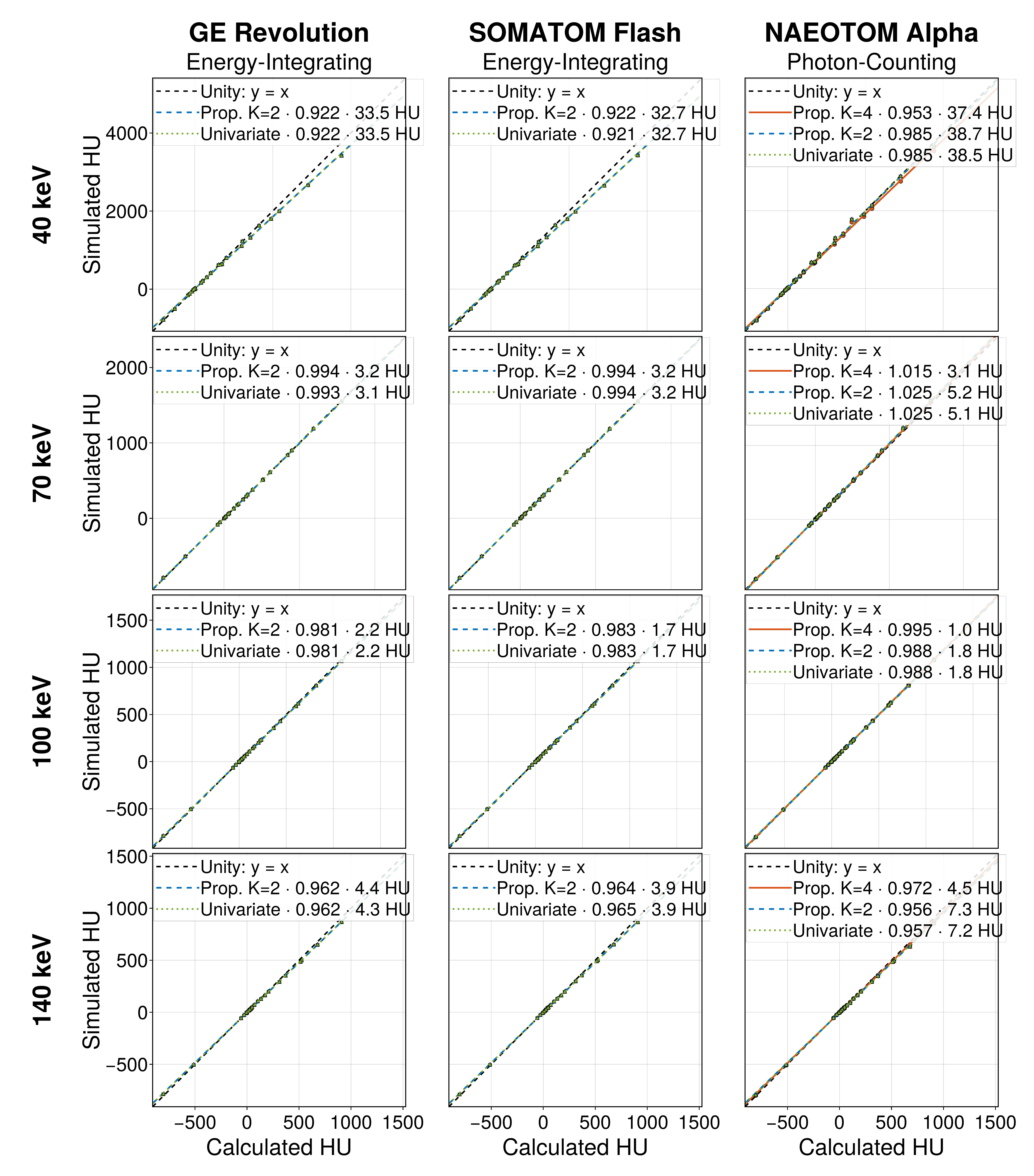}
\caption{Regression of simulated HU against calculated reference HU, with
the inserts of the Gammex 472 and tissue-characterization phantoms pooled in
each panel. Columns correspond
to the three acquisition configurations and rows to virtual monoenergetic energies of
40, 70, 100, and 140~keV. Results are shown for each decomposition method
supported by the corresponding acquisition. Each legend entry gives the
method (Prop., proposed), the regression slope, and the residual error{,
the root-mean-square residual about the fitted line in HU ($n=32$ inserts)}. {Each row
has its own axis range, equal on both axes, and horizontal tick values are shown
for the 140~keV row only.}}
\label{fig:regression}
\end{figure}

Tables~\ref{tab:rod_accuracy_gammex}
and~\ref{tab:rod_accuracy_tissue} summarize the CT-number accuracy for the individual
inserts of the photon-counting acquisition, which supports both the four- and
two-measurement decompositions, and Fig.~\ref{fig:regression} compares the
simulated and reference CT numbers for all three acquisitions. For the two-measurement {decompositions}, the proposed method
($K=2$) and the univariate method were in close agreement, with differences
within 0.001 in regression slope and 0.2~HU in residual error across the
three acquisition configurations. For the dual-energy acquisitions,
regression slopes ranged from 0.96 to 0.99 at 70~keV and above and were
approximately 0.92 at 40~keV.

For the photon-counting acquisition, increasing the number of measurements from $K=2$ to $K=4$
improved CT-number accuracy at 70, 100, and 140~keV. At 40~keV, however, the
MAE for the Gammex 472 phantom increased from 35.7 to
38.3~HU. This difference was strongly material dependent. For the iodine
inserts, the MAE decreased from 62.9 to 33.5~HU with $K=4$, whereas
for the calcium inserts it increased from 17.7 to 70.7~HU. A similar
material-dependent behavior was observed for the iodine and bone-equivalent
inserts of the tissue-characterization phantom.

The increased low-energy calcium error is consistent with limitations of
the water/iodine basis representation rather than failure of the numerical
optimization. {The attenuation of calcium is not an exact combination of those of the two
basis materials}, and this mismatch becomes more apparent near the iodine
K-edge at 33.17~keV. Retaining the individual photon-counting measurements
preserves this spectral information, whereas combining the lowest-energy
windows reduces its influence on the decomposition. Consistent with this
interpretation, every configuration under-read the high-density calcium
inserts at 40~keV. At
70~keV, the calcium error decreased from 22.7 to 15.0~HU when the number of measurements
increased from $K=2$ to $K=4$.

The additional spectral measurements provided a more consistent benefit in
image noise. Relative to $K=4$, reducing the photon-counting data to two
measurements increased the {mean within-insert SD} of the Gammex 472
phantom by
approximately 19\% at 40~keV and {8--12\%} at the higher VMI energies.
The regression residual error was also reduced with $K=4$, from 5.2 to 3.1~HU at
70~keV, from 1.8 to 1.0~HU at 100~keV, and from 7.3 to 4.5~HU at
140~keV. {The noise reduction is} consistent with the CRLB analysis, in which
retaining the additional spectral measurements reduced the statistical
uncertainty of the estimated basis coefficients.

\begin{table}[!htbp]
\centering
\caption{Insert-wise quantitative evaluation of the Gammex 472 phantom
for the photon-counting acquisition at 40, 70, 100, and 140~keV.
Calculated reference CT numbers (Calc.) are obtained from the known material
compositions, and physical scan reference CT numbers (Phys.) are measured from the
NAEOTOM Alpha virtual monoenergetic images. P4 and P2 denote the proposed method using four and
two spectral measurements, respectively, and Univ.\ denotes the
univariate method using the same two spectral measurements as P2.
The summary rows give, over the 16 inserts, the mean absolute error (MAE) and maximum absolute error (Max.\ error) of
$\Delta\mathrm{HU}_{\mathrm{ref}}$ (Eq.~\eqref{eq:hu_error}) against the
calculated and physical scan references, and the mean within-insert standard deviation (SD).}
\label{tab:rod_accuracy_gammex}
\scriptsize
\setlength{\tabcolsep}{2.4pt}
\resizebox{\textwidth}{!}{%
\begin{tabular}{lrrrrrrrrrrrrrrrrrrrr}
\toprule
& \multicolumn{5}{c}{40 keV} & \multicolumn{5}{c}{70 keV}
& \multicolumn{5}{c}{100 keV} & \multicolumn{5}{c}{140 keV}\\
\cmidrule(lr){2-6}\cmidrule(lr){7-11}\cmidrule(lr){12-16}\cmidrule(lr){17-21}
Insert & Calc. & Phys. & P4 & P2 & Univ. & Calc. & Phys. & P4 & P2 & Univ.
& Calc. & Phys. & P4 & P2 & Univ. & Calc. & Phys. & P4 & P2 & Univ.\\
\midrule
Pure water (1) & 0 & 1 & 8 & 14 & 14 & 0 & $-2$ & 1 & 1 & 1 & 0 & $-2$ & $-1$ & $-2$ & $-2$ & 0 & $-2$ & $-2$ & $-3$ & $-3$ \\
Pure water (2) & 0 & 6 & 8 & 10 & 11 & 0 & $-3$ & 3 & 3 & 3 & 0 & $-4$ & 2 & 2 & 2 & 0 & $-4$ & 1 & 1 & 1 \\
Solid water ref.\ (1) & 16 & 8 & 21 & 23 & 23 & $-1$ & $-2$ & 1 & 1 & 1 & $-5$ & $-2$ & $-4$ & $-5$ & $-5$ & $-6$ & $-4$ & $-6$ & $-7$ & $-7$ \\
Solid water ref.\ (2) & 16 & 19 & 18 & 26 & 26 & $-1$ & $-3$ & 1 & 2 & 2 & $-5$ & $-6$ & $-3$ & $-5$ & $-5$ & $-6$ & $-8$ & $-5$ & $-7$ & $-7$ \\
Ca 50 mg/mL & 311 & 323 & 310 & 319 & 319 & 175 & 194 & 179 & 180 & 180 & 145 & 162 & 145 & 144 & 144 & 134 & 149 & 131 & 129 & 129 \\
Ca 100 mg/mL & 673 & 659 & 644 & 664 & 664 & 313 & 340 & 320 & 322 & 322 & 235 & 259 & 236 & 233 & 233 & 206 & 225 & 201 & 197 & 197 \\
Ca 200 mg/mL & 1427 & 1309 & 1362 & 1409 & 1409 & 612 & 623 & 625 & 631 & 631 & 436 & 448 & 434 & 429 & 429 & 370 & 375 & 355 & 346 & 346 \\
Ca 300 mg/mL & 2166 & 1944 & 2056 & 2132 & 2131 & 899 & 906 & 921 & 932 & 932 & 627 & 638 & 626 & 620 & 620 & 525 & 528 & 505 & 492 & 492 \\
Ca 400 mg/mL & 2904 & 2561 & 2757 & 2885 & 2883 & 1187 & 1182 & 1215 & 1234 & 1234 & 817 & 824 & 815 & 805 & 805 & 679 & 677 & 650 & 629 & 629 \\
I 2.0 mg/mL & 165 & 160 & 192 & 206 & 206 & 55 & 48 & 59 & 61 & 61 & 26 & 21 & 25 & 24 & 24 & 14 & 10 & 11 & 8 & 8 \\
I 2.5 mg/mL & 207 & 209 & 216 & 228 & 228 & 68 & 68 & 71 & 72 & 72 & 31 & 31 & 33 & 32 & 32 & 16 & 16 & 17 & 15 & 15 \\
I 5.0 mg/mL & 408 & 427 & 427 & 438 & 438 & 130 & 131 & 134 & 136 & 136 & 57 & 55 & 58 & 58 & 58 & 27 & 24 & 26 & 25 & 25 \\
I 7.5 mg/mL & 609 & 629 & 646 & 671 & 671 & 192 & 191 & 200 & 203 & 203 & 82 & 78 & 83 & 82 & 82 & 38 & 32 & 36 & 32 & 32 \\
I 10.0 mg/mL & 810 & 815 & 853 & 891 & 891 & 253 & 257 & 262 & 269 & 269 & 108 & 116 & 108 & 107 & 107 & 49 & 57 & 45 & 40 & 40 \\
I 15.0 mg/mL & 1220 & 1181 & 1261 & 1297 & 1297 & 380 & 376 & 388 & 394 & 394 & 160 & 172 & 161 & 159 & 159 & 71 & 86 & 67 & 62 & 63 \\
I 20.0 mg/mL & 1647 & 1626 & 1706 & 1776 & 1775 & 518 & 515 & 529 & 541 & 540 & 223 & 229 & 223 & 220 & 220 & 103 & 111 & 98 & 88 & 88 \\
\addlinespace
MAE vs.\ Calculated & --- & --- & 38.3 & 35.7 & 35.7 & --- & --- & 8.1 & 12.7 & 12.6 & --- & --- & 1.1 & 2.7 & 2.7 & --- & --- & 5.9 & 11.0 & 10.9 \\
Max.\ error vs.\ Calculated & --- & --- & 147.4 & 129.0 & 127.8 & --- & --- & 28.4 & 47.1 & 46.9 & --- & --- & 2.5 & 12.3 & 12.2 & --- & --- & 28.6 & 50.0 & 49.8 \\
MAE vs.\ Physical & --- & --- & 41.7 & 70.1 & 69.7 & --- & --- & 9.9 & 14.0 & 14.0 & --- & --- & 7.9 & 9.5 & 9.5 & --- & --- & 10.9 & 15.0 & 14.9 \\
Max.\ error vs.\ Physical & --- & --- & 195.4 & 323.6 & 321.9 & --- & --- & 33.4 & 52.1 & 51.8 & --- & --- & 22.9 & 25.5 & 25.5 & --- & --- & 26.6 & 48.0 & 47.8 \\
Mean within-insert SD & --- & --- & 75.7 & 90.3 & 90.4 & --- & --- & 30.7 & 33.5 & 33.5 & --- & --- & 23.2 & 25.2 & 25.2 & --- & --- & 21.7 & 24.3 & 24.3 \\
\bottomrule
\end{tabular}}
\end{table}

\begin{table}[!htbp]
\centering
\caption{Tissue-characterization phantom, simulation only, on the same
photon-counting acquisition. Entries and columns are as in
Table~\ref{tab:rod_accuracy_gammex}{. This} phantom has no physical
counterpart, so there is no physical scan reference. The summary rows give, over
the 16 inserts, the mean absolute error (MAE) and maximum absolute error (Max.\ error) against the calculated reference and
the mean within-insert standard deviation (SD).}
\label{tab:rod_accuracy_tissue}
\scriptsize
\setlength{\tabcolsep}{2.6pt}
\resizebox{\textwidth}{!}{%
\begin{tabular}{lrrrrrrrrrrrrrrrr}
\toprule
& \multicolumn{4}{c}{40 keV} & \multicolumn{4}{c}{70 keV}
& \multicolumn{4}{c}{100 keV} & \multicolumn{4}{c}{140 keV}\\
\cmidrule(lr){2-5}\cmidrule(lr){6-9}\cmidrule(lr){10-13}\cmidrule(lr){14-17}
Insert & Calc. & P4 & P2 & Univ. & Calc. & P4 & P2 & Univ. & Calc. & P4 & P2 & Univ. & Calc. & P4 & P2 & Univ.\\
\midrule
Pure water & 0 & 12 & 13 & 13 & 0 & 2 & 3 & 3 & 0 & 0 & 0 & 0 & 0 & $-1$ & $-1$ & $-1$ \\
Solid water ref. & 16 & 29 & 31 & 31 & $-1$ & 3 & 3 & 3 & $-5$ & $-3$ & $-4$ & $-4$ & $-6$ & $-6$ & $-7$ & $-7$ \\
Lung LN-450 & $-805$ & $-818$ & $-824$ & $-824$ & $-805$ & $-804$ & $-805$ & $-805$ & $-805$ & $-801$ & $-800$ & $-800$ & $-805$ & $-799$ & $-798$ & $-798$ \\
Lung LN-300 & $-513$ & $-516$ & $-520$ & $-520$ & $-512$ & $-511$ & $-512$ & $-512$ & $-512$ & $-510$ & $-510$ & $-510$ & $-512$ & $-509$ & $-509$ & $-509$ \\
Adipose & $-176$ & $-163$ & $-159$ & $-159$ & $-84$ & $-84$ & $-84$ & $-84$ & $-65$ & $-63$ & $-64$ & $-64$ & $-57$ & $-55$ & $-56$ & $-56$ \\
Breast & $-132$ & $-117$ & $-120$ & $-120$ & $-52$ & $-51$ & $-51$ & $-51$ & $-36$ & $-34$ & $-33$ & $-33$ & $-29$ & $-27$ & $-26$ & $-26$ \\
Brain & $-70$ & $-51$ & $-47$ & $-47$ & 3 & 5 & 6 & 6 & 18 & 20 & 19 & 19 & 24 & 26 & 25 & 25 \\
Muscle & $-12$ & 1 & 7 & 7 & 18 & 21 & 22 & 22 & 24 & 26 & 25 & 25 & 27 & 28 & 27 & 27 \\
Liver & $-38$ & $-22$ & $-20$ & $-20$ & 25 & 27 & 27 & 27 & 38 & 40 & 40 & 40 & 43 & 45 & 45 & 45 \\
I 2.0 mg/mL & 165 & 172 & 173 & 173 & 55 & 56 & 56 & 56 & 26 & 26 & 26 & 26 & 14 & 13 & 13 & 13 \\
I 5.0 mg/mL & 408 & 414 & 420 & 420 & 130 & 131 & 132 & 132 & 57 & 57 & 57 & 57 & 27 & 27 & 26 & 26 \\
I 10.0 mg/mL & 810 & 818 & 828 & 827 & 253 & 255 & 256 & 256 & 108 & 109 & 108 & 108 & 49 & 49 & 46 & 47 \\
Trabecular bone & 701 & 665 & 678 & 678 & 293 & 299 & 300 & 300 & 206 & 204 & 202 & 202 & 173 & 165 & 161 & 161 \\
Bone CaCO\textsubscript{3} 300 & 1202 & 1135 & 1165 & 1165 & 514 & 523 & 525 & 525 & 366 & 364 & 359 & 359 & 311 & 298 & 291 & 291 \\
Bone CaCO\textsubscript{3} 500 & 1959 & 1849 & 1905 & 1906 & 843 & 857 & 864 & 864 & 604 & 600 & 593 & 593 & 514 & 494 & 482 & 482 \\
Cortical bone & 3763 & 3538 & 3654 & 3654 & 1559 & 1588 & 1604 & 1604 & 1087 & 1082 & 1072 & 1072 & 909 & 873 & 853 & 853 \\
\addlinespace
MAE vs.\ Calculated & --- & 36.0 & 25.1 & 25.0 & --- & 5.0 & 6.7 & 6.7 & --- & 2.1 & 3.3 & 3.3 & --- & 6.1 & 9.0 & 9.0 \\
Max.\ error vs.\ Calculated & --- & 225.2 & 109.1 & 109.4 & --- & 28.8 & 44.8 & 44.8 & --- & 5.0 & 15.0 & 15.0 & --- & 35.8 & 56.5 & 56.5 \\
Mean within-insert SD & --- & 49.9 & 57.4 & 57.5 & --- & 20.6 & 22.7 & 22.8 & --- & 16.2 & 18.0 & 18.0 & --- & 15.5 & 17.5 & 17.5 \\
\bottomrule
\end{tabular}}
\end{table}

\subsection{Comparison with physical scan measurements}
\label{sec:results_clinical}

\begin{figure}[!htbp]
\centering
\includegraphics[width=\textwidth]{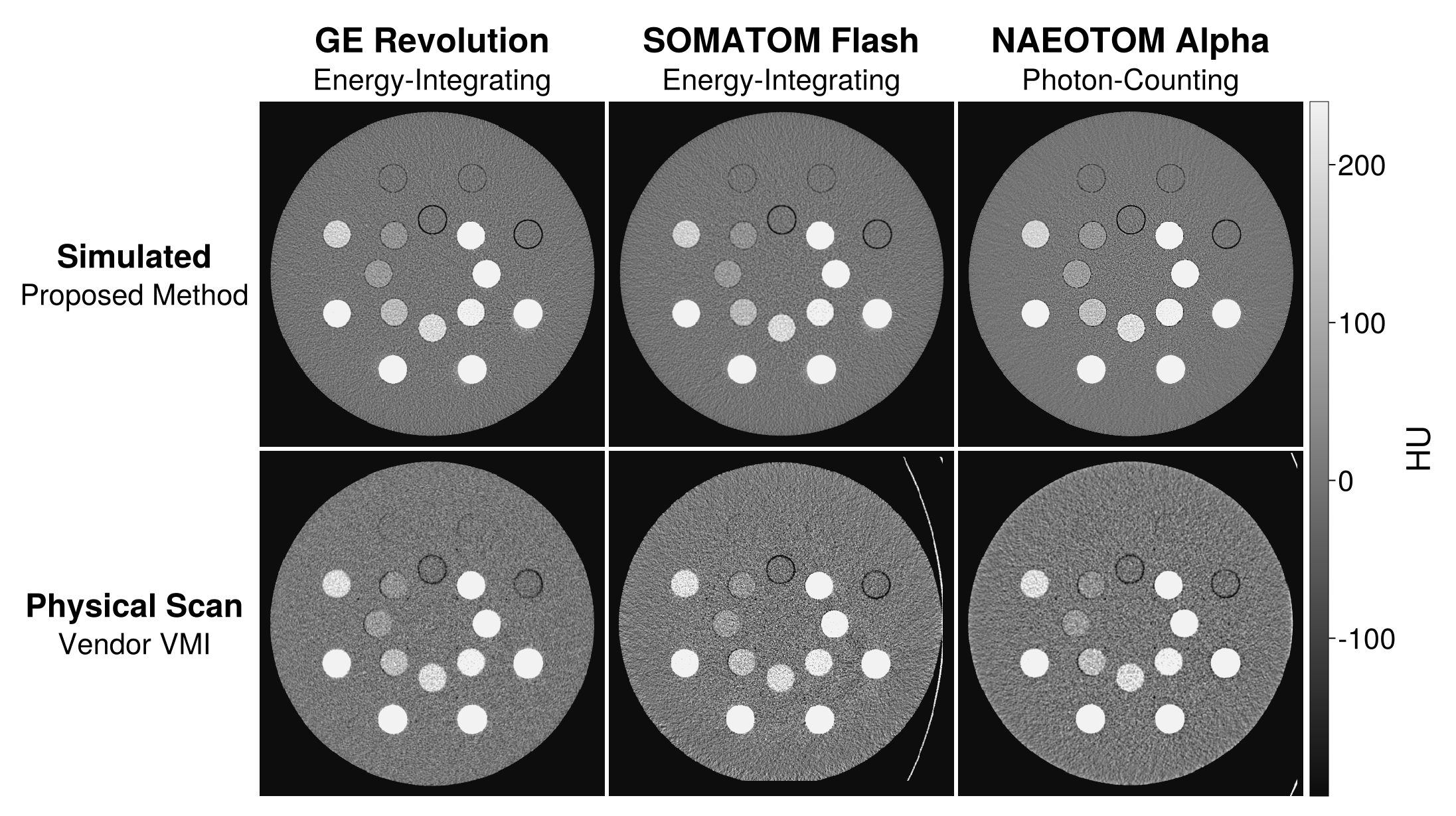}
\caption{Simulated and physical scan virtual monoenergetic images of the Gammex 472 phantom at
70~keV. Columns correspond to the three systems. The upper row shows the
simulation decomposed with the proposed method ($K=2$ for the GE Revolution
Apex Elite and SOMATOM Flash, $K=4$ for the NAEOTOM Alpha), and the lower row
shows the vendor's virtual monoenergetic image from the physical scan. Each image is cropped to the
phantom and shown in the orientation of the simulated phantom{. The} arc at the
edge of the SOMATOM Flash physical image is the boundary of its 33~cm
dual-energy field of view. All images use the same $[-200,240]$~HU window.}
\label{fig:physical_vmis}
\end{figure}

\begin{figure}[!htbp]
\centering
\includegraphics[width=\textwidth]{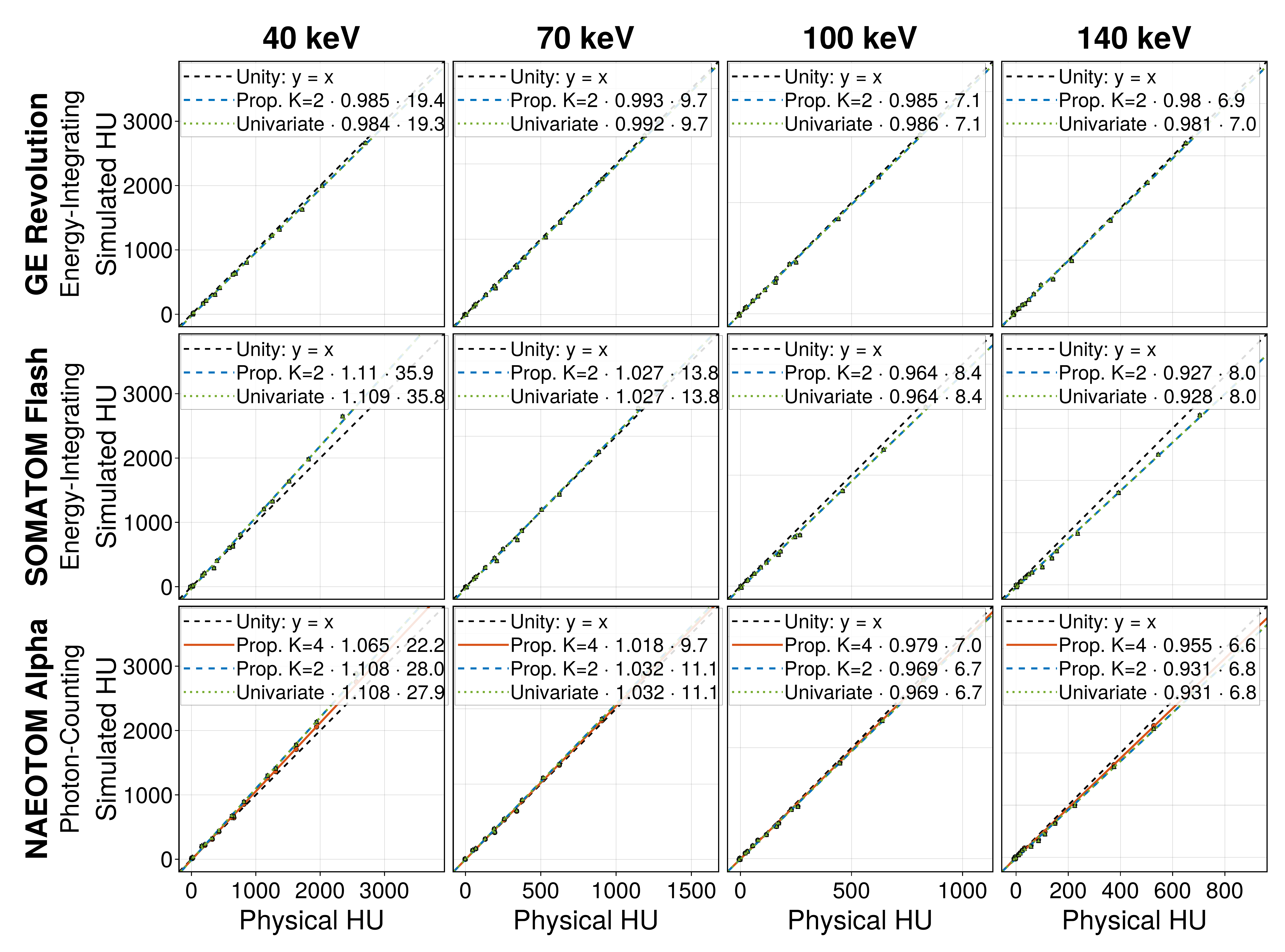}
\caption{Regression of simulated HU against physical scan HU for the Gammex 472
phantom. Rows correspond to the three acquisition configurations and
columns to virtual monoenergetic energies of 40, 70, 100, and 140~keV. Results are shown for
each decomposition method supported by the corresponding acquisition.
Each legend entry gives the method (Prop., proposed), the regression slope,
and the residual error in HU {($n=16$ inserts)}, and the dashed black line indicates unity. {Each column has its own axis
range, equal on both axes, and vertical tick values are shown for the 40~keV
column only.}}
\label{fig:clinical_regression}
\end{figure}

\begin{figure}[!htbp]
\centering
\includegraphics[width=\textwidth]{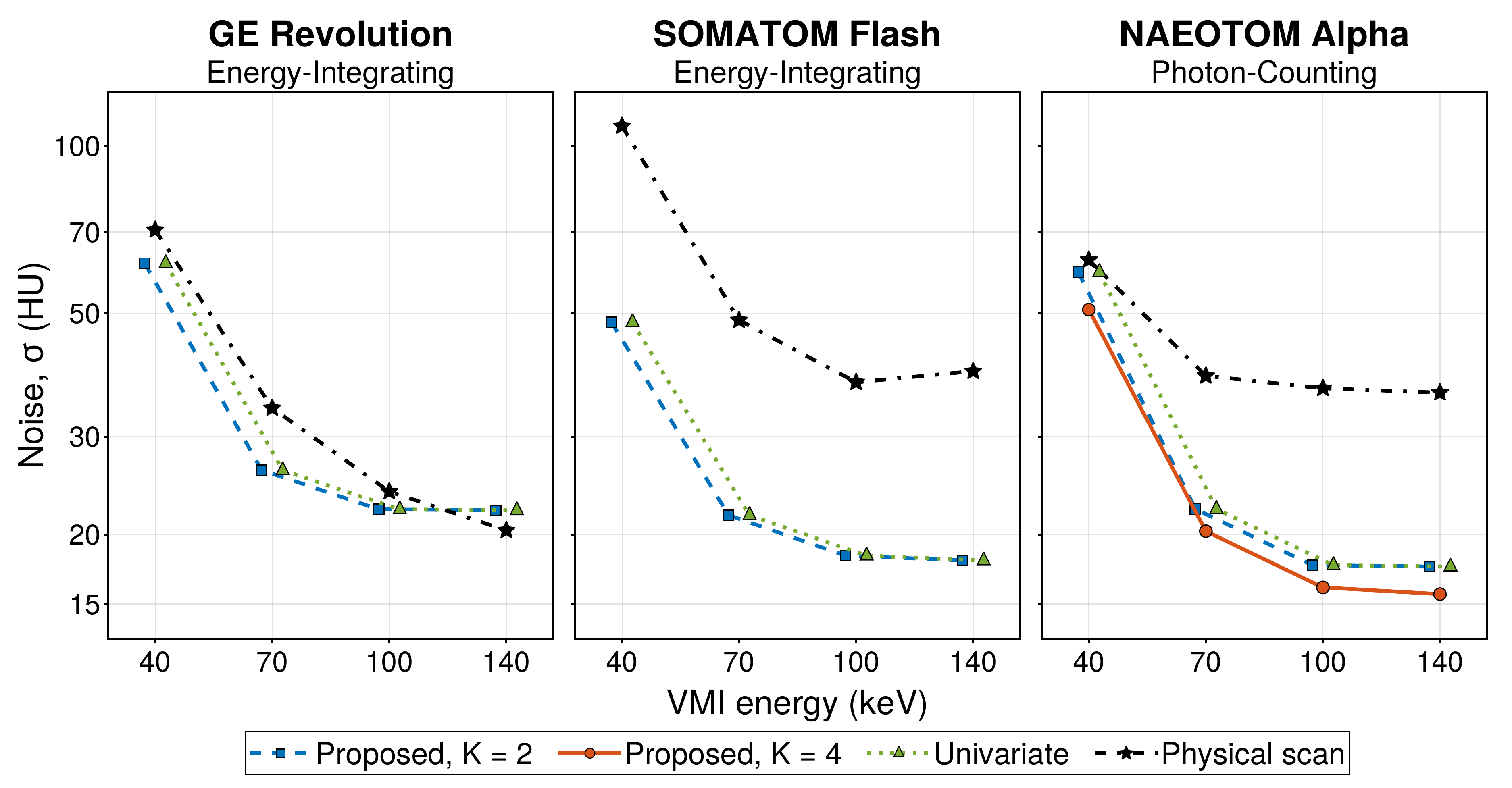}
\caption{HU standard deviation in a uniform solid-water region as a function of
virtual monoenergetic energy. Panels correspond to the three acquisition configurations and
show the proposed and univariate decomposition methods together with the
corresponding physical scan measurements. The vertical axis is shown on a
logarithmic scale.}
\label{fig:noise}
\end{figure}

The simulated Gammex 472 results were further compared with physical scans
of the same phantom on the three physical systems. Figure~\ref{fig:physical_vmis}
shows the simulated and physical scan VMIs side by side at 70~keV. Insert
contrast was visually similar, whereas the noise texture differed because
the physical scan VMIs include vendor-specific decomposition and noise
reduction. The NAEOTOM Alpha physical scan
HU values are reported in Table~\ref{tab:rod_accuracy_gammex}, and the corresponding
regression results are shown in Fig.~\ref{fig:clinical_regression}. Only
CT-number accuracy was compared directly because the physical scan VMIs include
manufacturer-specific decomposition and noise-reduction processing. The
physical scan noise measurements in Fig.~\ref{fig:noise} are therefore shown
for reference rather than as a direct comparison with the simulated noise.

For the photon-counting acquisition, the $K=4$ implementation showed closer
agreement with the physical scan measurements than $K=2$ across the evaluated
VMI energies. The largest difference occurred at 40~keV, where the
MAE relative to the physical scan measurements decreased from 70.1 to
41.7~HU. For the two dual-energy configurations, the MAEs
obtained with the proposed and univariate methods were within
0.8~HU of each other.

For the rapid kVp-switching acquisition, the MAE between the
simulated and physical scan measurements was approximately 6--11~HU at 70~keV
and above. For the dual-source acquisition, the corresponding difference
was approximately 11~HU at 70~keV and increased to 22~HU at 140~keV,
primarily because the calcium and iodine inserts had higher HU values in the
physical scan images than in the simulation. These differences include the effects of
scanner-specific spectral calibration, decomposition, and subsequent
image processing and therefore represent agreement between the simulated
and physical scan imaging pipelines rather than an independent measure of
decomposition accuracy.

At 40~keV, the trend relative to the physical scan measurements differed from
that obtained using the {calculated} reference values. In particular,
agreement with the physical scan images does not necessarily imply lower error
relative to the known material composition because the physical scan VMIs are
themselves generated through a vendor-specific material decomposition.
The {calculated} attenuation values therefore provide the primary reference
for quantitative accuracy, while the physical scan comparison evaluates the
agreement of the {simulation} with measurements from the physical
systems.

\subsection{Numerical sensitivity}
\label{sec:results_ablation}

Numerical sensitivity was evaluated using the Siemens NAEOTOM Alpha
photon-counting configuration. Table~\ref{tab:ablation} summarizes the
effect of independently varying each numerical parameter while holding the
remaining parameters at their default values. The method was stable over
a broad range of settings, with substantial changes occurring primarily
when the parameter bounds or update magnitudes became too restrictive.

The largest change occurred when the upper water bound was reduced to
25~g\,cm$^{-2}$, below the maximum water-equivalent path length of
approximately 33~g\,cm$^{-2}$. This caused 72.8\% of the rays to reach the
bound and changed insert HU by approximately 1450~HU. Reducing the
maximum iodine update to 0.01~g\,cm$^{-2}$ changed insert HU by
111~HU because the permitted update was insufficient to traverse the
required parameter range within the available outer iterations. Reducing
the iodine bounds produced a smaller change of 25~HU, with 0.22\% of
the rays reaching the bound. Extending the lower iodine bound slightly
below zero avoided boundary solutions associated with noisy measurements{.}
{Setting} the lower bound to zero increased the fraction of bound-limited
rays to approximately 13\%.

The remaining parameters of the proposed estimator produced changes of less than
0.04~HU over the evaluated ranges. In particular, the inner iteration
limit, Fisher condition limit, and air gate had negligible effect on
the reconstructed CT numbers. These results indicate that the selected
default parameters provide stable numerical behavior, with appreciable
degradation occurring only when the feasible parameter range or permitted
update size is substantially restricted.

\begin{table}[!htbp]
\centering
\caption{Numerical sensitivity for the Siemens NAEOTOM Alpha
photon-counting configuration. Each parameter was varied independently over
the specified range while the remaining parameters were held fixed.
Reported quantities are the maximum change in insert HU at each virtual monoenergetic energy,
the maximum change in water-region $\sigma_{\mathrm{HU}}$, and the maximum
fraction of rays reaching a parameter bound. {Results are for the Gammex 472
phantom, with $K=4$ for the proposed estimator and $K=2$ for the univariate
method, whose bracketed searches have no parameter bounds (---).} For the univariate method, $y$ is
the variable of its outer search in the notation of
Cong et al.~\cite{Cong2022}, not a measured count.}
\label{tab:ablation}
\small
\setlength{\tabcolsep}{4.5pt}
\begin{tabular}{lrcrrrrrrr}
\toprule
& \multicolumn{3}{c}{Range swept} & \multicolumn{4}{c}{$\max|\Delta\mathrm{HU}|$} & $\max\Delta\sigma$ & At bound\\
\cmidrule(lr){2-4}\cmidrule(lr){5-8}
Control & From & & To & 40 keV & 70 keV & 100 keV & 140 keV & (HU) & (\%)\\
\midrule
\multicolumn{10}{l}{\textit{Proposed estimator}}\\
$\Delta_A$ (max iodine step) & $0.01$ & to & $0.20$ & 111 & 13.4 & 12.0 & 22.4 & 0.24 & 0.0\\
$\Delta_C$ (max water step)  & $1.0$ & to & $20.0$ & $<0.01$ & $<0.01$ & $<0.01$ & $<0.01$ & $<0.01$ & 0.0\\
$\tau$ (parameter tolerance) & $10^{-3}$ & to & $10^{-6}$ & 0.03 & $<0.01$ & $<0.01$ & $<0.01$ & $<0.01$ & 0.0\\
$N_{\mathrm{outer}}$ & $8$ & to & $64$ & $<0.01$ & $<0.01$ & $<0.01$ & $<0.01$ & $<0.01$ & 0.0\\
$N_{\mathrm{inner}}$ & $6$ & to & $48$ & 0 & 0 & 0 & 0 & 0 & 0.0\\
Iodine bounds & $\times 0.5$ & to & $\times 2$ & 25.0 & 3.35 & 2.28 & 4.59 & 0.09 & 0.22\\
Water bounds & $\times 0.5$ & to & $\times 2$ & 1450 & 315 & 487 & 558 & 307 & 72.8\\
Fisher condition limit & $10^{6}$ & to & $10^{10}$ & 0 & 0 & 0 & 0 & 0 & 0.0\\
Air gate & $0$ & to & $10^{-3}$ & 0 & 0 & 0 & 0 & 0 & 0.0\\
\midrule
\multicolumn{10}{l}{\textit{Univariate method}}\\
Newton iterations & $4$ & to & $48$ & 5.75 & 1.06 & 0.16 & 0.66 & 0.07 & ---\\
Newton tolerance & $10^{-4}$ & to & $10^{-9}$ & $<0.01$ & $<0.01$ & $<0.01$ & $<0.01$ & $<0.01$ & ---\\
$y$ max factor & $0.90$ & to & $0.999$ & $<0.01$ & $<0.01$ & $<0.01$ & $<0.01$ & $<0.01$ & ---\\
$y$ max cap & $10^{5}$ & to & $10^{9}$ & 0 & 0 & 0 & 0 & 0 & ---\\
\bottomrule
\end{tabular}
\end{table}

\subsection{Effect of photon-counting corrections}
\label{sec:results_pcct}

In the detector response (Fig.~\ref{fig:pcct_corrections}), charge
sharing and fluorescence escape lower every window's peak response and give it
a long tail of photons from higher energies. Uncorrected pile-up moves counts
upward: window 4 also counted photons well below its threshold, and on the
central air ray, 30--31\% of the counts left
windows 1--3 while window 4 gained 68\%. After pile-up correction the response
is the Monte Carlo response that the decomposition models. Scatter, which adds
photons rather than moving them between windows, raised windows 1--3 by
12--17\% and window 4 by 6\% on the central ray through the phantom without
correction, and every window was within 2\% after correction.

\begin{figure}[!htbp]
\centering
\includegraphics[width=\textwidth]{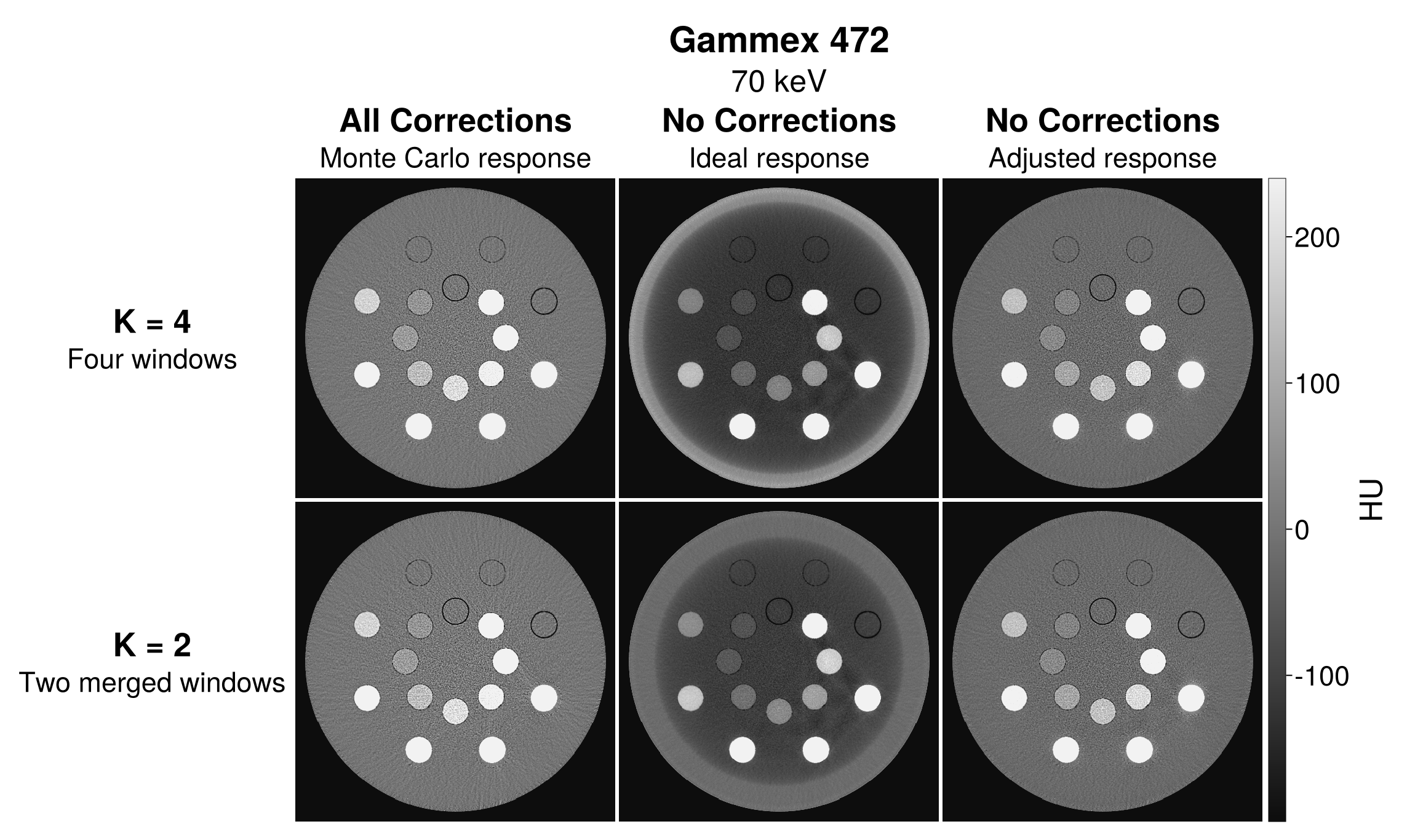}
\caption{Virtual monoenergetic images of the Gammex 472 phantom at 70~keV from
the photon-counting acquisition, using the four energy windows ($K=4$) and the
same two merged windows ($K=2$). Left: all corrections, with the Monte Carlo
detector response. Middle: no corrections, decomposed with an ideal response.
Right: no corrections, decomposed with the response adjusted to include
pile-up. All images use the same $[-200,240]$~HU window.}
\label{fig:pcct_corrections_vmi}
\end{figure}

Table~\ref{tab:pcct_corrections} and Fig.~\ref{fig:pcct_corrections_vmi}
compare three decompositions. With no corrections and an ideal response, the
assumed response (Fig.~\ref{fig:pcct_corrections}, top) differed from the one
that produced the counts (middle), and the estimator could not fit the
measurements: 84\% ($K=4$) and 77\% ($K=2$) of the rays reached the iodine
bound, the MAE rose from 38.3 to 806.0~HU at 40~keV and from 8.1 to 185.3~HU
at 70~keV ($K=4$), and the images showed strong cupping. Their low
within-insert SD reflects estimates held at the bound, not lower noise. The
ideal response is a worst case rather than a practical alternative: clinical
systems calibrate their response, which captures the fixed detector effects. With
the same uncorrected counts but the response adjusted to include pile-up, the
decomposition recovered: the MAE was 6.5 and
9.9~HU at 100 and 140~keV, against 1.1 and 5.9~HU with every correction. The
remaining error was largest at 40~keV (242.4 against 38.3~HU with every
correction) and is attributed to the scatter, which was corrected in the
reference configuration, and so in all earlier results, but which no response
can represent. Scatter raised windows 1--3 more than window 4, and the 40~keV
VMI amplifies errors in the basis estimates the most. In this configuration $K=4$ gave a
lower MAE than $K=2$ at every energy, 242.4 against 281.3~HU at 40~keV and 9.9
against 19.9~HU at 140~keV, and lower noise. In a physical detector, pile-up
depends on the count rate of each ray, so folding it into the response would
require a rate-dependent model.

\begin{table}[!htbp]
\centering
\caption{Mean absolute error (MAE) against the calculated CT numbers and mean
within-insert standard deviation (SD), both in HU, over the 16 inserts of the
Gammex 472 phantom, for the photon-counting acquisition. All corrections:
pile-up and scatter corrected, with the Monte Carlo detector response. No
corrections: pile-up and scatter left in the counts, decomposed with an ideal
response or with the response adjusted to include pile-up. $K=4$ uses the four
energy windows and $K=2$ the same two merged windows as in
Table~\ref{tab:rod_accuracy_gammex}.}
\label{tab:pcct_corrections}
\small
\setlength{\tabcolsep}{3pt}
\resizebox{\textwidth}{!}{%
\begin{tabular}{lllrrrrrrrr}
\toprule
& & & \multicolumn{4}{c}{MAE (HU)} & \multicolumn{4}{c}{SD (HU)}\\
\cmidrule(lr){4-7}\cmidrule(lr){8-11}
Corrections & Response & $K$ & 40 keV & 70 keV & 100 keV & 140 keV & 40 keV & 70 keV & 100 keV & 140 keV\\
\midrule
All  & Monte Carlo & 4 & 38.3 & 8.1 & 1.1 & 5.9 & 75.7 & 30.7 & 23.2 & 21.7\\
     &             & 2 & 35.7 & 12.7 & 2.7 & 11.0 & 90.3 & 33.5 & 25.2 & 24.3\\
None & Ideal       & 4 & 806.0 & 185.3 & 53.3 & 76.5 & 13.3 & 13.1 & 13.2 & 13.3\\
     &             & 2 & 793.9 & 167.2 & 46.9 & 78.4 & 12.4 & 12.1 & 12.3 & 12.4\\
None & Adjusted    & 4 & 242.4 & 46.0 & 6.5 & 9.9 & 60.9 & 25.7 & 20.6 & 19.9\\
     &             & 2 & 281.3 & 48.4 & 7.0 & 19.9 & 65.8 & 26.8 & 22.0 & 21.8\\
\bottomrule
\end{tabular}}
\end{table}

\section{Discussion}
\label{sec:discussion}

This work demonstrates that the univariate approach to projection-domain
material decomposition, in which the two-variable inversion is reduced to
one-dimensional searches~\cite{Cong2022}, can be generalized in two ways: from two
spectral measurements to an arbitrary number, and from a formulation
written for measurements separated by the x-ray source spectrum to
measurements separated by any part of their spectral response, including the
detector. The {method}
was evaluated using rapid kVp-switching, dual-source, and photon-counting
acquisitions without changing the underlying decomposition formulation.
{Maximum-likelihood estimators} are available for {photon-counting data
with several energy windows}{~\cite{Roessl2007,Schlomka2008}, including fast table-based
estimators that approach the CRLB~\cite{Alvarez2011}. Iterative searches over both basis line integrals jointly,
however,} require many forward-model evaluations per ray and can depend on
initialization and algorithmic parameters~\cite{Rajbhandary2017,Cong2022}. The profile
minimization retains the likelihood as the objective while reducing the
search to two one-dimensional problems, and the estimates approached the CRLB {for a representative ray}
on all three acquisitions with {small iodine bias (at most 0.04 standard
deviations)} (Table~\ref{tab:solver}). These
results show that the proposed formulation is not tied to a particular
scanner architecture and provides a common approach for projection-domain
material decomposition across different spectral CT systems. Because the
same estimator was applied to all three simulated acquisitions, the
differences between them {are not due to different
decomposition algorithms}. This matters for virtual imaging trials, which
often require several imaging systems to be simulated with reasonable
fidelity~\cite{Abadi2020}, typically without access to the vendors' own
{spectra, calibrations,} and processing. A single decomposition method that serves every system,
and whose VMIs agreed with the vendor VMIs of the physical scans to within {an}
MAE of 6--22~HU at 70~keV and above, provides a common step for such
trials. Because the estimator is released in the same open-source package as
the simulator, that step needs no separate implementation.

For $K=2$, the proposed {method} was benchmarked against the univariate
method. {The two methods agreed closely in} reconstructed VMIs, CT-number accuracy, regression performance, and
image noise across the three acquisition configurations. This agreement
provides a reference for the $K=2$ implementation and, more importantly,
allows the same formulation to be extended naturally when additional
spectral measurements are available. The replacement of
the polynomial relation by the profile minimization appeared in the line
integrals rather than in the images: on noiseless dual-energy data, the
largest iodine error of the univariate method was 10 to 15 times that of the
proposed method (Table~\ref{tab:solver}). The proposed method required {1.8} to 2.2
times the decomposition time of the univariate method but remained below one
second per sinogram on the GPU.

The advantage of this extension was demonstrated using the photon-counting
acquisition, where all four spectral measurements could be retained and used
directly for decomposition. Taguchi~\cite{Taguchi2024} showed{, for a simulated photon-counting detector,} that the CRLB on
decomposition noise decreases as energy {windows} are added. Here, combining the
four NAEOTOM Alpha windows into two raised the iodine {CRLB standard deviation} by 23\%, and
because each estimator came within a few percent of its own bound, the
difference arises {primarily} from the information lost when the four measurements are
reduced to two. Unlike a {two-measurement} implementation,
which requires the available measurements to be combined before
decomposition, the generalized {method} allows each spectral measurement to
contribute directly to the estimation of the basis components. The improved
quantitative performance with $K=4$ {at 70~keV and above} {indicates} that the additional
measurements contain useful information that can be retained by the proposed
formulation. Because the spectral responses are specified for each ray, the
same formulation also accepts detector responses that vary between pixels, a
recognized difficulty in photon-counting decomposition~\cite{Naidu2026},
provided those responses are known.

A similar benefit was observed in image noise, which was lower with $K=4$ at
every VMI energy on both phantoms (Tables~\ref{tab:rod_accuracy_gammex}
and~\ref{tab:rod_accuracy_tissue}, Fig.~\ref{fig:noise}). {Although the absolute noise levels depend on the acquisition,
reconstruction, and denoising, both decompositions used the same counts and
identical processing. The relative reduction therefore largely isolates the effect of
the number of measurements.} This reduction is
particularly useful for low-energy VMIs, where noise amplification following
material decomposition is more pronounced~\cite{McCollough2015}. These results
support the use of multi-measurement decomposition as the number of
energy-resolved measurements available from photon-counting CT continues to
increase.

The benefit of $K=4$ was also consistent between the Gammex 472 and
tissue-characterization phantoms{, and it was largest for the iodine- and
calcium-bearing inserts}. The exception was
calcium at 40~keV, where the error increased with $K=4$. Because the solver
recovered the noiseless {iodine} line integrals to within
$2.7\times10^{-6}$~g\,cm$^{-2}$ (Table~\ref{tab:solver}), this error {is attributed to}
the water/iodine basis {rather than} the {solver}. The physical NAEOTOM
Alpha 40~keV images also read the densest calcium inserts below their
calculated values (Table~\ref{tab:rod_accuracy_gammex}), so the deviation is
not specific to the proposed estimator.

Unlike regularized and one-step decompositions, whose results depend on tuned
weights~\cite{Long2014,Ducros2017,Mory2018}, the proposed {method} has only
numerical controls, and the numerical sensitivity analysis {showed that its
results were insensitive to them unless a bound or step limit was too
restrictive}.
Over the ranges evaluated in Table~\ref{tab:ablation}, the iteration limits,
the Fisher condition limit, and the convergence tolerance changed the
reconstructed CT numbers by less than 0.04~HU, and every appreciable
degradation occurred when a bound or step limit was tightened. The results therefore indicate that the
defaults are a plateau rather than a tuned operating point, and that stable
performance is obtained over a broad range of settings provided the water bound
exceeds the longest path through the object and the iodine bounds leave room
for per-ray noise.

The photon-counting comparison shows that the decomposition tolerates the
detector's nonidealities when they are included in the response it models
(Section~\ref{sec:results_pcct}). With an ideal response and no corrections,
the estimator could not fit the data. With the same uncorrected counts and a
response that included pile-up, charge sharing and escape, it recovered to
within a few HU of the fully corrected result at 100 and 140~keV, and four
windows remained better than two. In this simulation, only scatter, which adds photons rather
than moving them between windows, had to be corrected in the data. This is
the same principle that makes the method independent of the acquisition
mechanism: the detector enters the decomposition only through $S_k(E)$.

This study has limitations that point to directions for future work. The
estimator was evaluated in simulation and was not applied to measured
projection data, and the physical scans were compared only through the
vendor VMIs, which include each manufacturer's own decomposition and
processing. Each system was also modeled from its reported acquisition
parameters, without access to the vendor's {spectra}, calibrations, or
processing, so the simulations {only approximate the
physical systems.} Because the simulated data were generated with the same
spectral model used for decomposition, the next step is to apply the
estimator to measured projections, with its spectral model calibrated to
each system, which will also test its sensitivity to spectral-model
mismatch. {The proposed method was also benchmarked against the
univariate method it generalizes, not against a joint two-dimensional
likelihood search.} {Four} further extensions follow from the formulation. First, the
rapid kVp-switching and dual-source acquisitions were simulated with
co-registered rays{. Because} the physical systems do not provide these
directly~\cite{McCollough2015,Maass2011}, applying the estimator to them
will require bringing their measurements onto common ray paths{, as
rapid kVp-switching systems do by interpolating between neighboring views}. Second, {electronic noise}, which the Poisson likelihood of
Eq.~\eqref{eq:poisson_model} does not model separately
(Section~\ref{sec:likelihood}), could be incorporated through a likelihood
with an additional Gaussian term~\cite{XuTsui2009}. Third, a water/iodine
basis cannot represent calcium exactly, which causes the 40~keV deviation
on the calcium-bearing inserts, and extending the {method} to a different
or larger basis~\cite{Tang2021} would address such materials. {Fourth, the
photon-counting detector modeled here was CdTe. Deep-silicon detectors, which
use edge-on silicon sensors with more energy windows and a response shaped
largely by Compton scattering in the sensor~\cite{Persson2014}, were not
modeled. Because the {method} accepts any response $S_k(E)$, the same
principles should hold, but this remains to be tested in simulation.}
{Finally, as} in every projection-domain decomposition, the basis sinograms carry the
noise amplification inherent to the inversion~\cite{Kalender1988,Lee2023}, and the quality of the
synthesized VMIs depends on the denoising chain that follows. Because each
ray is decomposed independently, the output remains compatible with
sinogram-domain filters such as the likelihood-based bilateral filter used
here~\cite{Lee2023,LeeKim2025}. Future work
will focus on {denoising methods} that exploit information
shared across the spectral measurements while preserving the spectral
information gained by increasing $K$.

\section{Conclusion}
\label{sec:conclusion}

This work presents a unified projection-domain material decomposition
{method} for spectral CT with an arbitrary number of measurements. The
{method} was evaluated across rapid kVp-switching, dual-source, and
photon-counting acquisitions, with the $K=2$ results showing close agreement
with the established univariate method. Using all four photon-counting
measurements reduced noise at every VMI energy and improved quantitative
accuracy at 70~keV and above, compared with two merged measurements, demonstrating the benefit of retaining
the available spectral information. These results provide a foundation for
projection-domain material decomposition as spectral CT systems move toward
a greater number of energy-resolved measurements. Because one method serves
every acquisition and is built into an open-source CT simulator, {it} is also suited to virtual imaging trials
that model several proprietary spectral CT systems, whose own decomposition
algorithms are not available, within one consistent setting.

\section*{Acknowledgments}
The authors have no conflicts of interest to declare. AI tools were used to assist in preparing the manuscript and its associated code. The authors reviewed every part of the text and code and take full responsibility for the content.

\section*{Funding}
This work was supported in part by the National Heart, Lung, and Blood Institute of the National Institutes of Health under award R01HL170149.

\section*{Data and code availability}
The proposed estimator and the univariate method are included in the open-source \texttt{BasisSimulator.jl} package from version 0.15.0 (\url{https://github.com/MolloiLab/BasisSimulator.jl}). The simulations in this paper were performed with version 0.14.0 of the same package. The simulated acquisitions can be {regenerated} with the package, and the Gammex 472 phantom is provided in its documentation.

\end{document}